# Ethics Whitepaper: Whitepaper on Ethical Research into Large Language Models

Eddie L. Ungless
University of Edinburgh

Nikolas Vitsakis
Heriot-Watt University

Zeerak Talat
University of Edinburgh

James Garforth
University of Edinburgh

Björn Ross
University of Edinburgh

Arno Onken
University of Edinburgh

Atoosa Kasirzadeh
University of Edinburgh

Alexandra Birch*
University of Edinburgh

*This whitepaper offers an overview of the ethical considerations surrounding research into or with large language models (LLMs). As LLMs become more integrated into widely used applications, their societal impact increases, bringing important ethical questions to the forefront. With a growing body of work examining the ethical development, deployment, and use of LLMs, this whitepaper provides a comprehensive and practical guide to best practices, designed to help those in research and in industry to uphold the highest ethical standards in their work.*

## 1. Introduction

### *Motivation and intended audience of this whitepaper*

As large language models (LLMs) grow increasingly powerful, their advancements in natural language understanding and generation are impressive (Min et al. 2023). However, mitigating the risks they present remains a complex challenge, and categorising these risks is a crucial aspect of ethical research related to LLMs (Weidinger et al. 2022). Key concerns include the potential to perpetuate and even amplify existing biases present in training data (Gallegos et al. 2024), the challenges in safeguarding user privacy (Yao et al. 2024), hallucination or incorrect responses (Abercrombie et al. 2023; Xu, Jain, and Kankanhalli 2024), malicious use of their powerful capabilities (Cuthbertson 2023), and infringement of copyright (Lucchi 2023). Given that many of these ethical challenges remain unresolved, it is essential for those involved in developing LLMs and LLM-based applications to consider potential harms, particularly as these models see broader adoption.

Several frameworks have already been developed to address AI ethics and safety. For example The U.S. National Institute of Standards and Technology (NIST) has a AI Risk Management Framework (RMF) [1], which provides broad guidelines for managing

* 10 Crichton Street, EH89AB, UK, a.birch@ed.ac.uk
1 https://www.nist.gov/itl/ai-risk-management-framework

AI-related risks. NIST has also recently released a document outlining specific risks and recommended actions for Generative AI [2]. While widely adopted, the NIST guidelines are voluntary. In contrast, the EU AI Act [3] represents a legally binding regulatory framework designed to ensure the safe and ethical use of AI within the European Union. It emphasises transparency, human oversight, and the prevention of discriminatory outcomes, with the goal of protecting fundamental rights and promoting trustworthy AI.

The NIST AI RMF and EU AI Act are broad, focusing on AI deployment and risk management across industries. There are other frameworks which are more research-focused, guiding ethical considerations in academic AI work. For example the Conference on Neural Information Processing Systems (NeurIPS) Ethics Guidelines [4] evaluates AI research for ethical concerns as part of the paper submission process. A similar effort from the Association of Computational Linguistics (ACL) has created an Ethics Checklist[5] which guides authors in addressing ethical implications, including limitations, and correct treatment of human annotators.

Despite there being a number of frameworks for the ethical development of AI, we believe that there is still a need for a practical whitepaper focused on the needs of a practitioner working with LLMs. This whitepaper presents insight and pointers to the most relevant ethical research, as it relates to each of the steps in the project lifecycle. It provides more detail than the guidelines of NeurIPS and ACL, but is more "digestible" and directly applicable to research with LLMs than the NIST frameworks or the EU AI act. We hope this whitepaper will prove valuable to all practitioners, whether you are looking for succinct best practice recommendations, a directory of relevant literature, or an introduction to some of the controversies in the field.

## 2. Overview

This document is structured around a (simplified) project lifecycle, depicted in Figure 1. Our aim for this whitepaper is for it to be used as a reference guide throughout a project, rather than for post-hoc reflection. We begin in Section 3 by outlining the importance of ethics and discussing themes relevant to the entire development life cycle. In Section 4 we consider best practice for data collection and sharing. Next, in Section 5 we discuss ethical aspects of data preparation, such as cleaning and labelling. We then turn towards model development in Section 6, focusing on questions of model design, addressing social biases and model alignment. Following a common develop-then-test structure, we then consider ethical issues related to performance and harm evaluation in Section 7. Finally in Section 8, we examine the questions of ethics that arise in deployment contexts. You should of course adapt the order you consult these sections to your needs e.g. those finetuning an existing LLM may wish to consult Section 6 on Model Development (which has advice for model selection) before Section 4 on Data Compilation (for guidance on compiling finetuning data). However, we recommend every practitioner starts with Section 3, as this has vital guidance for all projects.

At the end of each Section we give key resources, namely concrete do's and don'ts for ethical research, relevant to that stage on the project, and tools to guide ethical work.

2 https://nvlpubs.nist.gov/nistpubs/ai/NIST.AI.600-1.pdf
3 https://digital-strategy.ec.europa.eu/en/policies/regulatory-framework-ai
4 https://neurips.cc/public/EthicsGuidelines
5 https://aclrollingreview.org/responsibleNLPresearch/

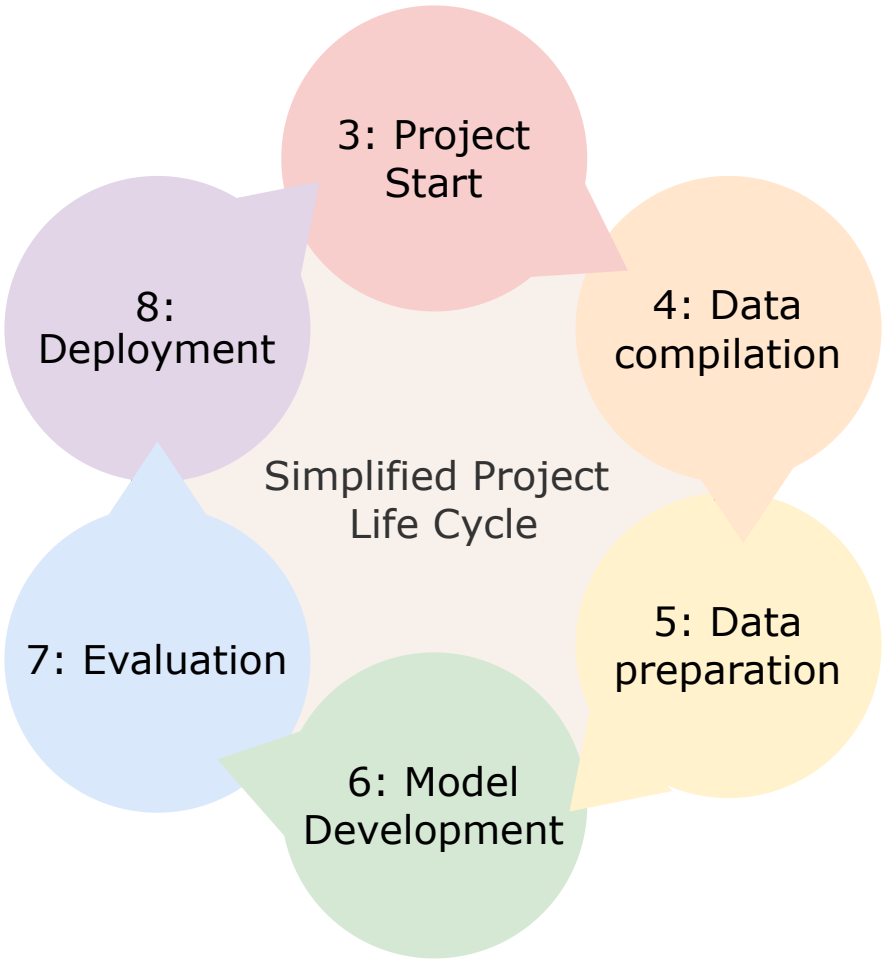

**Figure 1**
Diagram showing simplified project lifecycle that forms the structure of this paper.

## 3. Project Start

### *A quick overview of the broad-reaching ethical implications of LLMs*

The social risk of generative AI, and LLMs can have wide-reaching effects from representational harms to safety concerns, which has been widely recognised Weidinger et al. (see e.g., 2021); Bender et al. (see e.g., 2021); Uzun (see e.g., 2023); Wei and Zhou (see e.g., 2022). This recognition has given rise to a large number of efforts seeking to evaluate their risks and actualised harms, each effort presenting its own limitations (Solaiman et al. 2024; Goldfarb-Tarrant et al. 2023; Blodgett et al. 2020). Nevertheless, efforts towards developing technologies that minimise harms, in particular to marginalised communities, are vital. Best efforts require considering a wide range of topics and questions that must be adapted to each individual application and deployment context. In this Section we explain why ethics is relevant to all practitioners (Section 3.1), then highlight resources to aid in the initial discovery process (Section 3.2). We also lay out best practice that will be valuable to all those working with language technologies, namely related to working with stakeholders, and environmental considerations (Section 3.3 and Section 3.4).

### 3.1 Who needs ethics?

#### *Everyone needs ethics*

As computer science becomes pervasive in modern lives, so too does it become intertwined with the experience of those lives. Decisions made by researchers and developers compound together to influence every aspect of the technical systems which ultimately govern how we all live. This is often at a scale, or level of complexity, which make it impossible to seek clear resolutions when outcomes are harmful (Van de Poel 2020; Kasirzadeh 2021; Miller 2021; Birhane et al. 2022; Santurkar et al. 2023; Pistilli et al. 2024).

Technical artefacts are inherently political (Winner 1980), because they further entrench certain kinds of power e.g. marginalised peoples' data is often used without consent or compensation (see Section 4.2); technology typically works best for language varieties associated with whiteness (Blodgett and O'Connor 2017); benchmarks are published which are biased against minorities (Buolamwini and Gebru 2018). Unfortunately, the training and work cultures of computer scientists often condition us to believe we are beyond politics (Malazita and Resetar 2019), because the "objective" or abstractive nature of our work seems to absolve us of having to consider issues of our technologies in the world (Talat et al. 2021) – when dealing with code and numbers it becomes easier to forget about the real humans who are impacted by our design choices. LLMs are no exception (Leidner and Plachouras 2017), though their recent rise in prevalence has made their ethical dimensions more salient (and more vital to address).

Technical work should be considered interdisciplinary by its very nature, as it requires an understanding of the physical and social systems that technical artefacts must interact with in order to achieve their function. Experts exist in all of these other areas of study, as well as their intersections, but very often our lack of appreciation for their expertise, or lack of shared language, impede us from seeking them out. This is especially true for expertise in the social sciences and philosophy (Raji, Scheuerman, and Amironesei 2021; Inie and Derczynski 2021; Danks 2022).

There is a tendency to assume that social and ethical issues are someone else's problem (Widder and Nafus 2023), but this is not the case! If you do not reflect on your design decisions as you make them then you are complicit in the avoidable consequences of those decisions (Talat et al. 2021). The decision to follow a code of ethics (McNamara, Smith, and Murphy-Hill 2018) or employ a pre-packaged ethical toolkit, does not immediately solve the problem because these decisions require a level of ethical reflection to be effective (Wong, Madaio, and Merrill 2023).

### 3.2 Laying the Groundwork

*Resources to ensure ethical issues considered from the beginning*

It is important to think about ethics from the very beginning, in order to be able to question all aspects of the project, including if specific tasks should even be undertaken. One way of doing this is by using ethics sheets (Mohammad 2022), which are sets of questions to ask and answer before starting an AI project. It includes questions like "Why should we automate this task?", and "How can the automated system be abused?". An alternative is using the Assessment List for Trustworthy Artificial Intelligence (ALTAI)[6], which is a tool that helps business and organisations to self-assess the trustworthiness of their AI systems under development. The European Commission appointed a group of experts to provide advice on its artificial intelligence strategy and they translated these requirements into a detailed Assessment List, taking into account feedback from a six month long piloting process within the European AI community. You could use question sets such as these to ensure ethical considerations are present from the start of your project.

6 https://ec.europa.eu/newsroom/dae/document.cfm?doc_id=68342

Regulated industries such as aerospace, medicine and finance have critical safety issues to address, and a primary way these have been addressed is using auditable processes throughout a project. Audits are tools for interrogating complex processes, to determine whether they comply with company policy and industry standards or regulations (Liu et al. 2012). Raji et al. (2020) introduce a framework for algorithmic auditing that supports artificial intelligence system development, which is intended to contribute to closing the gap between principles and practice. They map out the process of implementing an internal audit for use during the development process (as opposed to post-hoc audits), which you can adopt to ensure responsibility for upholding ethical values is clearly assigned. A formal process such as this can help by raising awareness, assigning responsibility, and improving consistency in both procedures and outcomes (Leidner and Plachouras 2017). At the very least, your organisation should establish an ethics review board to evaluate new products, services, or research plans (Raji et al. 2020).

Foreseeing the downstream effects of deploying an AI system is challenging and requires grappling with complex contexts that even seemingly simple AI technologies may interact with. The ideation tools discussed above will help in anticipating possible effects. You will also benefit from reading about likely risks, for example in Solaiman et al. (2024) (see also Section 8.1). Further resources include Buçinca et al. (2023) who propose AHA! (Anticipating Harms of AI) which combines crowd-workers and LLMs to surface a wide variety of potential harms.

Weisz et al. (2023) outline design principles for developing generative AI applications, grounded in the concept of generative variability, which highlight that generative AI systems produce outputs which can vary in quality and character. The design principles emphasize designing for multiple, imperfect outputs. Their final design principle is proactively mitigating potential harms such as producing inappropriate content or displacing human workers. You can use these design principles to help you make thoughtful decisions when developing innovative LLM applications.

Finally, exploring ethics in a global context has the added challenge of balancing between the valuing of cultural diversity and the respect for human rights. Reid et al. (2021) develop a framework for collaboration that respects the variety of people and cultures that might be involved in research and encourages researchers to consider their perspectives from ideation right through to the legacy left by the project.

### 3.3 Stakeholders

***Best practice for thinking about and working with stakeholders***

Given the vast amounts of training data required and the wide-reaching applications of LLMs, every project will have many stakeholders e.g. those who provide the data (Havens et al. 2020), end-users of the application (Yang, Li, and Wei 2023), or those a model will be used *on*, who are often given limited power to influence design decisions e.g. migrants (Nalbandian 2022). A vital early step is identifying key direct stakeholders and establishing the best ways to work with them in order to build systems that are widely beneficial, and this will be highly context dependent (Sloane et al. 2022). The ideation toolkits detailed above will help you to identify stakeholders, and can be used alongside existing taxonomies e.g. (Lewis et al. 2020; Langer et al. 2021; Bird, Ungless, and Kasirzadeh 2023; Havens et al. 2020, i.a.). Crucially, stakeholders should be identified before development, so they can (if they wish) be involved in co-production, or object to proposed technologies (Munn 2022). Kawakami et al. (2024) present a toolkit for early stage deliberation with stakeholders with question prompts, while Caselli et al. (2021) provide 9 guiding principles for effective participatory design (which involves mutual learning between designer and participant) in the context of NLP research. Input from stakeholders is vital for constructing Value Scenarios, a methodology which predicts the likely impact of a proposed technology (Nathan, Klasnja, and Friedman 2007), as in Haroutunian (2022)'s paper on low-resourced machine translation. Havens et al. (2020) invite us to work collaboratively with stakeholders to understand who or what is included in our research, and who is excluded, while Jurgens, Hemphill, and Chandrasekharan (2019) explain how working with affected communities enables the establishment of research norms that are context-informed and sensitive to the needs of the community (e.g., establishment and use of appropriate language (Jurgens, Hemphill, and Chandrasekharan 2019)).

You must consider power relations between stakeholders (Havens et al. 2020), including between yourself and the stakeholders. Those of marginalised genders or ethnicities may not be represented in your research team (West, Whittaker, and Crawford 2019), which can influence researcher-stakeholder relations (e.g. lack of cultural knowledge, greater power imbalance (Madaio et al. 2022; Fukuda-Parr and Gibbons 2021; Haque et al. 2024)). Reflexive considerations about a researcher's own power are rare in computer science research (Ovalle et al. 2023b; Devinney, Björklund, and Björklund 2022) but can help establish the limitations of your work (Liang, Munson, and Kientz 2021; Liang 2021). This is particularly important when reaching out to marginalised communities, as these relationships can be even unintentionally exploitative. For example marginalised communities are often not fairly compensated for their participation (Sloane et al. 2022), or are only invited to partake in feedback but not co-production (Sloane et al. 2022; Ungless, Ross, and Lauscher 2023).

When working on technologies for indigenous and endangered languages, sensitive stakeholder collaboration is particularly important (Bird 2020; Liu et al. 2022; Mahelona et al. 2023). Work on stakeholder engagement in NLP can learn much from the Indigenous Data Sovereignty movement (Sloane et al. 2022), which we return to in Section 4.2.

### 3.4 Energy Consumption

#### *Consider the ethics of the environment*

Throughout the life cycle of a project, you should consider the energy consumption of your model, which relates to data sourcing practices, model design, choice of hardware, and use at production. Strubell, Ganesh, and McCallum (2019) suggest that model development likely contributes a "substantial proportion of the... emissions attributed to many NLP researchers". Strubell, Ganesh, and McCallum (2019) call for more research on computationally efficient hardware and algorithms, and the standardised calculation and reporting of finetuning cost-benefit assessments, so researchers can select efficient models to finetune (models that are responsive to finetuning). Similar recommendations are made by Henderson et al. (2020), who also provide a framework for tracking energy, compute and carbon impacts. Patterson et al. (2022) provide best practice for reducing the carbon footprint, including the development of sparse over dense model architectures and the use of cloud computing that relies on renewable energy sources. Bannour et al. (2021) provide a taxonomy of tools available to measure the impact of NLP technologies.

Sasha Luccioni and colleagues have in particular championed the accurate reporting of the carbon emissions of ML systems including LLMs (Luccioni, Viguier, and Ligozat 2023; Luccioni and Hernandez-Garcia 2023; Wang et al. 2023; Luccioni, Jernite, and Strubell 2024; Dodge et al. 2022; Lacoste et al. 2019). They find "better performance is not generally achieved by using more energy" across a range of NLP tasks (Luccioni and Hernandez-Garcia 2023). They provide a Python package for tracking the carbon impact of a given codebase (Courty et al. 2024), and a procedure to measure the energy consumption of LLMs during finetuning (Wang et al. 2023). Wang et al. (2023) note one way to reduce energy consumption at fine-tuning and inference is to use compression techniques, such as pruning and distillation, the additional cost of distilling quickly offset by these models being more energy efficient. However you should note that compression techniques can impact fairness as we discuss further in Section section 6.1; clearly, compromises must be made. Luccioni, Jernite, and Strubell (2024) also highlight that huge multi-purpose models have far greater cost at inference, so if classification is required it would be more energy efficient to use a task-specific model than rely on prompting.

### 3.5 Key Resources

Do's and Don'ts

- **Do** engage with affected communities from the beginning - **don't** just ask for their feedback
- **Do** allow for flexibility in project direction as informed by stakeholder input - **don't** assume what communities want and need
- **Do** consider the power relations between stakeholders – **don't** forget about the relationships with yourself
- **Do** engage with ethics review boards to ensure oversight, or set one up if necessary - **don't** assume because its computer science that moral and political values are out of scope
- **Do** create an internal audit procedure to ensure ethical processes are developed and followed - **don't** just leave it to a post-hoc review
- **Do** consider use of compressed models and cloud resources to minimise energy impact - **don't** assume you need energy intensive models for the best performance

Useful Tool(kit)s:

- Ethics sheets to discover harms and mitigation strategies – Mohammad (2022)
- The Assessment List for Trustworthy Artificial Intelligence (ALTAI) [7]
- Internal audit framework to ensure that ethical processes are implemented and followed – Raji et al. (2020)
- Value Scenarios framework to identify likely impact of technology – Nathan, Klasnja, and Friedman (2007)

7 https://ec.europa.eu/newsroom/dae/document.cfm?doc_id=68342

- Guiding principles for effective participatory design – Caselli et al. (2021)
- Best practice for reducing carbon footprint during training – Patterson et al. (2022)
- Taxonomy of tools available to measure environmental impact of NLP technologies – Bannour et al. (2021)
- Software package to estimate carbon dioxide required to execute Python codebase – `https://github.com/mlco2/codecarbon`

## 4. Data compilation

***Overview of ethically significant impacts of data compilation decisions***

### 4.1 Compiling own "raw" data

***Issues with typical practices in data compilation and guidance on best practice.***

When collecting data, we have to make deliberate choices about what data to include, or not; the ethical ramifications of these choices should be carefully considered. Design biases in data set compilation often stem from the creator's positionality (Talat et al. 2021; Santy et al. 2023): people with different lived experiences will make different choices. Therefore, a useful first step may be to consider your own positionality, what choices you are making and *why*, and to document this - a task that is much easier said than done! However, here we can borrow methods from qualitative fields, such as the social sciences. In particular, the concept of *reflexivity* can help researchers identify how your own positionality and subjectivity provides a particular view into your work (Jamieson, Govaart, and Pownall 2023).[8]

Reflexive research practices can also help to critically examine how we see the data we work with. There is the pervasive notion of data as a natural resource, ready to be exploited. As Benjamin (2021) notes, this metaphor is harmful. Data is created by people, and to exploit the data can mean exploiting the people who created it. As Gitelman and Jackson (2013) point out: "raw data is an oxymoron." For example, the texts we use to train LLMs have been created by someone for a particular purpose, and we reuse them for another. The supposedly "raw" data has actually already undergone various transformations: from someone's thought to written text, in a particular context and time, and then further to "data" that is stripped of its context.

A common source of "raw" data for LLM training is texts that are publicly available on the internet (see e.g. OpenAI 2024). The ethics of compiling data for training LLMs is therefore closely linked to the ethics of web crawling which have long been discussed in the context of web search (Gold and Latonero 2017). There are clear differences, however, between web search and LLM training, and the social norms established for search should be revisited: we may well conclude that there is data that is ethical to index for search purposes but unethical to copy for LLM training purposes. As for the "dark web" that is not indexed by search engines, ethical issues abound when it comes to scraping this data for the purpose of training LLMs. For example, online discussions

8 For an example of a reflexive practive in the context of NLP research, see Talat (2021).

on social media platforms such as review sites may be appropriate to index, using data from discussion boards where people explicitly seek to hide their identity and avoid association to their person may not be ethical, or even wanted as the data may contain artefacts that are undesirable to encode in LLMs.

Data scraped from the internet is likely to contain sensitive personal information (Mieskes 2017) (we return to this in Section 5.1) and over-represent certain populations, languages, and speakers (Dunn 2020). This can lead to issues of unfairness; while unfairness can enter a model at many different stages (Suresh and Guttag 2021), training data selection is certainly one of them. For an overview of the types of bias this can result in, see Navigli, Conia, and Ross (2023). For another, broader perspective on issues in data set creation, see Paullada et al. (2020).

Of course there are many other potential approaches to obtain data, such as licensing it from rights holders. The prospect of holding data from multiple rights holders about millions of data subjects who may be unaware of the data processing, while operating across a variety of jurisdictions, bring up complex questions of big data governance that Jernite et al. (2022) have attempted to address. The sheer size and diversity of these data sets makes it practically impossible to fully understand them, and filter out unwanted data; Luccioni and Viviano (2021) propose a number of research directions to address this issue. These rights holders are unlikely to spontaneously undermine their own business models, however, and so those of us licensing from them should be sure to be holding them to high standards.

### 4.2 Consent and Safety

***Issues related to consent, safety and power when compiling data from people.***

We follow Havens et al. (2020) and Bird, Ungless, and Kasirzadeh (2023) in distinguishing those who produce data from those who are represented in the data (data subjects) (e.g. a tweet might be about another person). Many jurisdictions afford legal protections to both: data producers are protected by copyright laws, which we discuss in Section 4.3, while data subjects (can) benefit from privacy legislation. However, different jurisdictions adopt different approaches, some offering far less protection. You should reflect on whether you are doing enough to protect both data producers and subjects, and consider extending protections beyond the extent required by law. Taking due ethical care, beyond what the law requires, is often the best way to build public affection and enthusiasm for new technologies, rather than alienating the communities we design for.

A major issue is lack of consent from data providers and subjects. People are often unaware their public data is being collected for LLMs (Kim et al. 2023) (as Scheuerman et al. (2023); Meng, Kekülluöğlu, and Vaniea (2021) discuss for other AI technology), or are given insufficient information to consent (Winograd 2022; Xiao 2020). Some large publicly available vision and language data sets have introduced processes for data producers and subjects to have data removed (Heikkilä 2022), for example in line with GDPR e.g. LAION.[9] This protection cannot help if the data has already been used to train a model, which may go on to reproduce training data, including Personally Identifiable Information (PII) (Lukas et al. 2023) and copyright material (Karamolegkou et al. 2023). You must respect the privacy rights of data producers and subjects (which

9 https://laion.ai/faq/

can be conflicting (Kekulluoglu, Kokciyan, and Yolum 2018)), rather than relying on the public to request their data be removed. You should use best practice such as transparent data collection, right to removal and respect of copyright to keep your data as clean as possible.

Best practice for social media data collection involves only storing post IDs, such that deleted posts will not be included in future uses. Even where social media platform policies mention content may be used for research, it is debatable whether this constitutes informed consent (Fiesler et al. 2024). Fiesler et al. (2024) recommend reaching out to community moderators on Reddit to understand privacy expectations, which is also relevant to forums and Facebook groups. Mancosu and Vegetti (2020)

Data from vulnerable populations or about sensitive topics must be especially carefully handled. Klassen and Fiesler (2022) emphasise the importance of cultural competence when using social media data from marginalised groups, to mitigate inadvertent harm. Data de-anonymsation can risk data subject safety (Rocher, Hendrickx, and de Montjoye 2019). The UK's Information Commissioner's Office provides guidelines based on GDPR which may be useful, for example by identifying the kinds of sensitive data that may need additional security and defining informed consent.[10] When gathering data, reflect on what data you need to accomplish the task, and what sensitive data should be excluded. Benton, Coppersmith, and Dredze (2017) provide guidelines for compiling social media data for studying health; their guidelines may be useful to those working with social media data across different sensitive tasks, for example, they advise to "separate annotations from user data".

Certain communities may be under-represented in your data, due to (a) lack of public online presence (b) small community size (c) problematic data compilation practices (Guyan 2021). This leads to biased model output (see section 6.2 and section 7.2). Markl (2022) provides case studies for evaluating speech data sets for what and who is missing, which can be easily extended to other modalities. When addressing under-representation, care must be taken not to cause harm. It may not be safe for certain communities to be compiled into labeled data sets e.g. queer populations (Ungless, Ross, and Lauscher 2023; Sigurgeirsson and Ungless 2024). Data compilation practices can be exploitative, for example when data producers are not given the knowledge required for informed consent, or when monetary incentives act as a form of coercion (Fussell 2019; Reid et al. 2021; Mahelona et al. 2023). When compiling data from marginalised populations, the Indigenous data sovereignty movement provides best practice (Walter et al. 2021). For example, giving marginalised populations agency in what data is collected about them, and taking into account the worldviews of the affected population.

### 4.3 Sharing and Using data

***Ethical considerations when sharing data or using shared data, including documentation.***

Training LLMs relies heavily on very large data sets composed of web-crawled data and other data sources. A useful position paper (Rogers, Baldwin, and Leins 2021) proposes a checklist for responsible data use and reuse. The legal and ethical principles of data collection are complicated by the sheer number of sources of data and jurisdictions.

10 https://ico.org.uk/for-organisations/advice-for-small-organisations/frequently-asked-questions/data-storage-sharing-and-security/#whatsecurity

Researchers need to follow legal restrictions, for example by respecting copyright, but this can vary depending on the country. In addition to copyright, data is frequently protected by "Terms of Service" for example by social media companies like X (formerly Twitter). These vary by site and can change over time. Ethical principles such as respecting privacy, allowing reproducibility and doing no harm can also be complicated to follow and even conflicting. How do we respect Twitter users' privacy by not saving their tweets, while allowing other researchers access to our data? The checklist proposed by Rogers, Baldwin, and Leins (2021) advances the discussion towards establishing a single standard for responsible data compilation.

Though we rely heavily on large data sets for training LLMs, there is a significant gap in our understanding of their content, including general statistics, quality, social factors, and the presence of evaluation data i.e. contamination. Elazar et al. (2024) introduces "What's In My Big Data?" (WIMBD), a platform designed to uncover and compare the contents of large text corpora, finding issues of low-quality content, sensitive data and benchmark contamination in ten corpora used in training popular language models, such as C4 and The Pile.

One of the biggest contributions researchers can make to the field is publishing new data sets. However, it is very important that future users of these data sets understand why and how they were created in order to use them correctly. Following work detailing how to describe data sets specifically for natural language processing (Bender and Friedman 2018), a more broadly applicable approach was proposed called datasheets (Gebru et al. 2020). Datasheets are structured documents that provide information about a data set to ensure its appropriate and ethical use, including details on its composition, intended uses, and limitations. This transparency helps users understand the data's context and potential harms.

### 4.4 Key Resources

Do's and Don'ts

- **Do** reflect on and document the decisions you make when collecting data - **don't** forget that *how* you collect data transforms it
- **Do** consider if it is ethical to scrape web content, even for content that is publicly available (e.g., by relying on frameworks of ethical data scraping such as Mancosu and Vegetti (2020)) - **don't** crawl content that website creators have indicated should not be crawled (e.g. via `robots.txt` files)
- **Do** consider the subjects of the data - **don't** just think about the rights of data producers
- **Do** respect copyright and privacy from the beginning - **don't** expect the public to do the work of requesting removal (but give them the option!)
- **Do** provide a datasheet for any data set you produce - **don't** forget to document intended use and limitations

Useful Tool(kit)s:

- Case study structure to identify who is missing from collected data – Markl (2022)

- Best practice from Indigenous data sovereignty movement – Walter et al. (2021)
- Checklist for responsible data collection and reuse - Rogers, Baldwin, and Leins (2021)
- API to explore content of popular massive data sets - Elazar et al. (2024)
- Guidelines to create datasheets - Gebru et al. (2020)

## 5. Data preparation

***Overview of ethically significant impacts of data preparation decisions***

### 5.1 “Cleaning” data

***Potential ethical issues and best practice for data cleaning e.g. filtering.***

It is vital that you conduct some “cleaning” of training data to minimise the amount of irrelevant (e.g. content from another language, when creating a monolingual model), offensive (e.g. hatespeech, misogyny, violence) or private (e.g. personally identifiable information, PII) content, which would limit model performance or result in harmful output. However, automatically filtering out “irrelevant” language data has been shown to introduce racial biases due to the association between “non-standard” languages and marginalised ethnic communities (for example African American Aligned English (Blodgett and O’Connor 2017)). Jurgens, Tsvetkov, and Jurafsky (2017) has proposed a methodology for language identification that allows for greater diversity.

Use of n-gram lists or machine learning based systems to detect and filter out harmful content can both introduce issues of bias (Anwar et al. 2024). Word lists may exclude reclaimed slurs and queer identity terms (Bender et al. 2021). Hate detection models such as Perspective API have been shown to exhibit bias against content by and about marginalised communities (Röttger et al. 2021), and have poor robustness (Calabrese et al. 2021). Further, filtering out explicitely offensive language may fail to identify subtle forms of implicit harm (Gadiraju et al. 2023; Talat et al. 2017), including “dog whistles”[11] that seem innocuous. What data is considered “dirty”, needing to be cleaned out, is highly subjective (Thylstrup and Talat 2020). It will benefit you to think of data harm as a spectrum rather than a binary category. As such it would be reasonable to “draw the line” at different “levels” of harm, so whatever decision you make should be justified. Further different kinds of harmful content will benefit from different cleaning techniques.

Hong et al. (2024) highlight how filtering practices can introduce and reinforce biases (in the context of text-image models) – which has also been demonstrated for language models (Welbl et al. 2021) – and provide recommendations for those conducting data filtering more broadly, including justifying your filtering design choices and evaluating chosen filters for bias. Anwar et al. (2024) propose relying on data balancing and rewriting in place of data filtering to minimise the unfair impact.

11 https://swu-union.org.uk/2023/01/part-4-dog-whistles-in-context-transphobia/

As Subramani et al. (2023) note, the topic of removing PII from training data for LLMs is “relatively under-explored”, which is particularly problematic given the tendency of LLMs to memorise and later reveal PII such as email addresses (Carlini et al. 2021). Subramani et al. (2023) define risk levels and categories of PI, and provide guidance on how to detect and remove “character based” PI (i.e. sequences of numbers or letters that uniquely identify a person such as a credit card number or email address). They find C4 and Pile, widely used crawled training data sets, containing tens of millions of instances of PI. They recommend use of out of the box PI detection tools such as Presidio to find and replace PI with anonymising tokens.

### 5.2 Choice of target labels

***Guidelines for defining target (prediction) label options for your data.***

Labelling can be thought of as the process of simplifying reality (as approximated in your data) into a format that is readable to an ML system. This can introduce a great number of ethical issues. The proxy may be an *over*-simplification (Mulvin 2021), for example treating offensive language as a binary between explicitely offensive and not offensive, which fails to account for implied offense (see ElSherief et al. 2021). When designing labels, we recommend that you make explicit what information you are aiming to record, and what information is lost through your choice of proxy.

Guerdan et al. (2023) discuss how proxy choice impacts outcomes in human-AI decision models, and their guidance is valuable to all supervised training paradigms. They invite researchers to ask themselves questions such as: does the proxy serve as a “satisfactory approximation” of the target variable (unobserved by the model)? They highlight risks to construct reliability through e.g. poor specification of the annotation protocol, as well as through annotator bias.

Harm may occur when target labels (and also metadata) are proxies for complex social constructs such as gender. Larson (2017) provide best practice when using gender as a variable (label) in NLP, and their advice can be applied to other sensitive categories: for example, you should make theories of gender (/identity) explicit, which might look like explaining that you consider gender to include “man”, “woman” and “nonbinary” categories. They have advice for handling metadata also. See also Savoldi et al. (2021); Dev et al. (2021) for clear discussions of the relationship between gender and language, which may be useful for designing labels. Working directly with affected communities to define labels allows for greater cultural sensitivity, as demonstrated by Maronikolakis et al. (2022) and Dev et al. (2024).

For some tasks, silver labelling (assignment of non-expert labels by some heuristic means) may be necessary for reasons of efficiency, but you should be mindful about poor quality of silver labels, as Lignos et al. (2022) discuss for multilingual corpora. Their recommendations to improve quality can also apply more broadly, for example advising “consider the potential negative consequences of releasing data sets known to be of low quality... they will likely be used for evaluation purposes".

### 5.3 Impact of Annotators

***Summary of the impact of annotators’ lived experiences on annotations.***

Research has shown that annotating behaviour can be influenced by various factors, such as education level (Al Kuwatly, Wich, and Groh 2020; Fortuna et al. 2022), cultural background (Smart et al. 2024), gender (Biester et al. 2022; Excell and Al Moubayed 2021), or even level of expertise in the annotating task (Talat 2016; Lopez Long, O'Neil, and Kübler 2021). For example, cultural background has shown to impact annotations of sarcasm and hate speech (Lee et al. 2024), and morality (Davani et al. 2024). These cultural differences are particualrly pressing given the tendency to rely on global majority labour to label Western data (Smart et al. 2024). While not often collected, differences between annotators' identities, attitudes, and beliefs have also been shown to influence annotations of toxicity (Sap et al. 2022), and hate speech (Feng et al. 2023). To counteract these issues, you should gather relevant information through validated psychometric scales or questionnaires (Abercrombie et al. 2024; Sap et al. 2022).

The crowdsourcing platform through which data is collected can also have an impact. For example, Amazon's Mechanical Turk (MTurk), has been criticised for producing low-quality data (Chmielewski and Kucker 2020; Fort et al. 2014). While some evidence exploring these effects suggests they are the outcome of researcher design decisions (Allahbakhsh et al. 2013; Daniel et al. 2018), other research suggests that quality of data can be platform specific (Chmielewski and Kucker 2020; Eyal et al. 2021; Kennedy et al. 2020). In any case, there is a clear need to incorporate the perspectives of the workers themselves when designing a crowdsource task, in order to ensure not only that the data produced is of high quality, but that the livelihood and wellbeing of annotators is explicitly addressed (Huang, Fleisig, and Klein 2023).

### 5.4 Handling Annotation Disagreement

***Best practice for handling annotation disagreement.***

The notion that every given example in a task can be paired with a single ground truth "gold label"(Basile 2020), is challenged by the simple fact that annotators disagree. While annotation disagreement can be an indication of task related difficulty (Talat et al. 2021), it can also be indicative of clashing perspectives, common in subjective tasks (Uma et al. 2021; Uma, Almanea, and Poesio 2022).

Research efforts that attempt to allow for the representation of multiple perspectives largely follow two main trends. **Disagreement based** methodologies advocate for the use of distributional labels which more accurately capture and reflect annotator agreement / disagreement throughout a data set (Mokhberian et al. 2024; Leonardelli et al. 2023; Uma, Almanea, and Poesio 2022; Uma et al. 2021). While preferable to use of singular gold labels, this approach still limits the amount of possible perspectives identifiable to two, there is no room for more nuanced minority perspectives to emerge. **Metadata based** methodologies encode annotator metadata with the aim to capture clear signal from groups which share metadata information labels (Rottger et al. 2022; Davani et al. 2023; Fleisig, Abebe, and Klein 2023; Gupta et al. 2023). These approaches assume that individuals who share similar metadata (e.g., same gender), will also annotate similarly; which might be why findings supportive of this methodology seem to be both data set (Lee, An, and Thorne 2023) and task (Welch et al. 2020) specific.

### 5.5 Rights of Crowdworkers

***Overview of issues related to crowdworkers' rights.***

A substantial amount of NLP research takes advantage of crowdworkers — workers on crowdsourcing platforms such as MTurk. For many years, the NLP field has been aware of ethical issues surrounding the use of crowdworkers (Fort, Adda, and Cohen 2011; Fort et al. 2014; Bederson and Quinn 2011). The most striking issue is the very low wages (often below $2 an hour), which is exaggerated given the additional "invisible" labour workers do (Hara et al. 2018; Toxtli, Suri, and Savage 2021), but this is just the start. Crowdwork often results in insecure employment lacking payment guarantees or basic workplace rights, such as unionization (cf Perrigo 2023)). Workers in this community have no access to redress channels for employer misconduct, unlike typical workers in many other developed nations, who can use lawsuits and complaints to government agencies.

More recent work (Shmueli et al. 2021) points out that the categorization of crowdworkers as human participants can be a gray area as regulations did not forsee this kind of worker platform. They assert that crowdworkers should indeed be considered human participants as we study not only their outputs but also their behaviours, and that payment does not absolve the researchers from needing to pursue ethical approval. They point out that ethics approval is extremely important as the annotation can lead to significant harms to the crowdworkers e.g. psychological harm; risks of unwittingly revealing private data, or of employing vulnerable people e.g. children or prisoners (Kaun and Stiernstedt 2020; Mason and Suri 2012).

### 5.6 Key Resources

Do's and Don'ts

- **Do** carefully reflect on *whose* data you are excluding when cleaning - **don't** rely on popular tools to give you fair results
- **Do** make explicit what information you are trying to record with your choice of proxy - **don't** forget that labels and proxies are simplifications
- **Do** work with affected communities to define labels and annotate your data – **don't** forget that harm is subjective, and a spectrum
- **Don't** release low quality data that may be repurposed for evaluation
- **Do** gather information about your annotators – **don't** assume annotators with similar identities will give similar annotations
- **Do** treat crowdworkers as human participants and follow best practice for human participant research, such as collecting informed consent; seek formal ethics (IRB) approval where applicable - **don't** assume that when using paid annotators you do not need to follow typical ethics procedures

Useful Tool(kit)s:

- Recommendations for those conducting data filtering – Hong et al. (2024)
- Taxonomy of personal information and best practice for privacy – Subramani et al. (2023)
- Guidance of selecting proxy labels – Guerdan et al. (2023)
- Best practice when using identity terms as labels – Larson (2017)
- Detailed overview of risks of using crowdworkers – Shmueli et al. (2021)

## 6. Model Development + Selection

***Overview of ethically significant impacts of model design and training decisions.***

### 6.1 Model Design

***What to consider when selecting a model or deciding parameters.***

There is a belief that ML models merely reflect or indeed amplify existing bias in the data set. However as pointed out by Hooker (2021), algorithms themselves are not impartial, and some design choices are better than others. For instance the choice of tokenisation can introduces unfairness between languages (Petrov et al. 2023) and between identities (Ovalle et al. 2023a), and the choice of encoding method can affect handling of non-standard dialects (Tan et al. 2020). Learning rate "can also disproportionately impact error rates on the long-tail of the data set", where potentially marginalised more identities are represented (Hooker 2021). Recognizing how subtle model design changes can impact harm leads to mitigation techniques that can be more effective than relying on more commonly used approaches like ensuring more representative data collection.

One of the main decisions is what size of model to deploy. Some earlier work (Rae et al. 2021) showed larger models are more likely to generate toxic responses when provided with toxic prompts, but they can also more accurately classify toxicity. Another impact of scale is that language models obtain capabilities that they can use for moral self-correction (Ganguli et al. 2023; Schick, Udupa, and Schütze 2021) which means that they have potentially more capacity to avoid producing harmful outputs if instructed to do so. These key capabilities are following instructions and learning concepts of harm like stereotyping, bias, and discrimination. It seems that more powerful models are a double edged sword but when used correctly can better mitigate some harms.

Choosing what compression to use also has ethical implications. Pruning, distillation and quantization are frequently used techniques. They can significantly worsen performance on underrepresented features, and this frequently aligns with fairness considerations (Ramesh et al. 2023; Hooker et al. 2020). However, there is some evidence that as the model is compressed, it forgets some toxic content and becomes less toxic (Xu and Hu 2022). In any case, you should test for bias and model safety on the final deployed model, not on a fuller version pre-compression.

If you are designing a specific application, built ontop of an LLM, then model cards will be indispensable for selecting the right LLM (Mitchell et al. 2019). Model cards accompany machine learning models and provide benchmarked evaluation in a variety of conditions, such as across different groups (e.g., race, geographic location), plus they disclose the intended use context and performance characteristics. Likewise it is good practice to create model cards for models that you train.

### 6.2 Fairness and Debiasing Techniques

***Limitations of techniques to mitigate harms of LLMs during training.***

There are a number of "model level interventions" (Kumar et al. 2022) that have been proposed to reduce LLM related harms. Models may be finetuned to reduce the risks of harm, including bias (Subramanian et al. 2021), for example Ung, Xu, and Boureau (2022) provide a finetuning data set for dialogue models to better handle communica-tion failures. Another method is use of reinforcement learning with human feedback (RLHF). RLHF is a technique popularised by OpenAI which involves learning a reward function based on human feedback on output, which is used to further refine the model (Huang et al. 2023a). Models can also be modified to remove specific information, for example through "finetuning to forget", and subspace ablation (Anwar et al. 2024). Other methods to reduce harm include modifying learning objectives (Guo, Yang, and Abbasi 2022); incorporating knowledge bases; and guardrails (see Section section 8.3). For an overview, see Kumar et al. (2022).

Model debiasing may have limited efficacy, however, while retraining LLMs and may be infeasible given available computing resources or impossible given limited access to a model. For example, it is not clear if upstream debiasing prevents bias after finetuning (Steed et al. 2022) (cf. Liang et al. 2020), and upstream intrinsic bias metrics, often used to evaluate debiasing techniques, have little relationship with downstream, extrinsic measures (Cao et al. 2022a; Goldfarb-Tarrant et al. 2021; Delobelle et al. 2022) (see Section 7.2). Further, debiasing can introduce new biases (Xu et al. 2021), and once popular debiasing techniques involving word vector similarity have been shown to be superficial (Gonen and Goldberg 2019). Issues with RLHF are covered below under the discussion of alignment. Model modification methods commonly rely on attributing model components to specific knowledge, but interpretability techniques often lack the required robustness (Anwar et al. 2024). Finally, the original purpose of the model places an "upper bound" on how ethical even a debiased model can be (Kasirzadeh and Klein 2021). Therefore we advise caution when relying on "debiased" models.

### 6.3 Alignment

***Alignment is unreliable and ill-defined.***

LLM value alignment is a complex and multifaceted challenge, beginning with the very definition of the concept itself. The notion of "alignment" is inherently ambiguous, meaning different things to different stakeholders: from adhering to human preferences and societal norms to upholding ethical principles or achieving specific operational outcomes (Hendrycks et al. 2020; Gabriel 2020; Kasirzadeh and Gabriel 2023; Kirk et al. 2023; Anwar et al. 2024). This lack of a universally accepted definition complicates efforts to "align" LLMs, as the target of alignment remains elusive and subject to interpretation. Furthermore, the object of alignment - be it human preferences, moral principles, or societal values - is itself a moving target (Anwar et al. 2024). These constructs are diverse, evolving, and often contradictory across cultures and individuals, raising questions about whose values should be prioritized and how to reconcile conflicting viewpoints (Askell et al. 2021; Sorensen et al. 2024; Pistilli et al. 2024).

The multi-stage nature of LLM development - from data collection and pre-training to fine-tuning and deployment - introduces multiple points where misalignment can occur or be introduced. Ensuring and guaranteeing alignment throughout this pipeline needs vigilance at each stage and mechanisms to prevent the dilution or distortion of alignment efforts as the model evolves.

### 6.4 Key Resources

Do's and Don'ts

- **Do** consider how subtle changes can improve performance for marginalised people - **don't** assume that all bias comes from data imbalance
- **Do** use and create model cards for documenting correct and indended uses of models - **don't** assume that a model will be reliable for all populations you might care about
- **Do** test for harm on the deployed model – **don't** test on larger versions before compression as harms can be amplified by this process
- **Do** explore techniques such as finetuning to mitigate harm - but **don't** forget this can introduce new harms, and may not be effective
- **Do** maintain vigilance to ensure alignment is maintained throughout a pipeline – **don't** assume there is one set of human values

Useful Tool(kit)s:

- Very clear explanation of how model design choices impact fairness – Hooker (2021)
- Templates to document ML models including intended use context – Mitchell et al. (2019)
- Overview of techniques to mitigate LLM harms – Kumar et al. (2022)

## 7. Evaluation

***Overview of ethically significant impacts of performance and harm evaluation.***

### 7.1 Performance Evaluation

***Ethical problems that arise during performance evaluation e.g. evaluation is not robust.***

Evaluation is a crucial component of developing high quality models which are unbiased and safe. Ideally we would perform comprehensive testing of a model's ability in a realistic setting but this is often infeasible. Therefore, the most commonly used paradigm for testing models is to calculate task-specific evaluative metrics on held-out test sets. This can lead to an overestimation of real-world performance, as the model may be deployed in diverse and unexpected situations (Ribeiro et al. 2020), and potentially further entrench existing ethical problems if the same biases are found in the training and test set.

*Cases where evaluation failed.* There are a number of papers demonstrating that evaluation of language generation is not reliable. Caglayan, Madhyastha, and Specia (2020) show that there are numerous failure cases for standard metrics like ROUGE, BLEU and BERTScore, for example they can prefer system output to human-authored text, because it is less varied. Researchers also find that these metrics are insensitive to rare words, making it hard to reliably evaluate rare linguistic phenomena. Furthermore, embedding based metrics (eg. BERTScore) can exhibit significantly higher bias than traditional metrics on attributes such as race, gender and socioeconomic status (Sun et al. 2022).

*Benchmarks.* Benchmarks attempt to address some of the above limitations by gathering together sets of evaluations. Benchmarks are heavily used as indicators of progress towards long-term goals such as achieving flexible and generalizable AI systems. For example both GLUE and MMLU are collections of language understanding tasks which have been widely used to mark progress in LLM research. However, benchmarks are inherently specific, finite and contextual (Raji et al. 2021) and not the broad measure of progress that they are purported to be.

More importantly benchmarks are not objective, as they themselves can be biased, for example because certain demographics are under represented (Buolamwini and Gebru 2018) or because mean results hide issues for particular intersecting demographics (Tatman 2017). A final danger is that as a benchmark becomes central to a field of research, it encourages uncritically chasing algorithmic improvement or "hill-climbing", and losing sight of the performance mismatch with the real world uses as a result. Further, there are some aspects of performance which are not suited at all to being tested by benchmarks, as LaCroix and Luccioni (2022) argue is the case for "ethicality", which cannot be measured without context.

*Solutions.* Although high quality and comprehensive benchmarks are part of the solution, they cannot solve all the problems previously discussed. We can also borrow ideas from software engineering, which has long experience with testing complex systems. We can perform “behavioural testing” or “black-box testing” which tests different capabilities of a system by validating the input-output behaviour, going beyond just testing accuracy on held-out data (Ribeiro et al. 2020). CHECKLIST (Ribeiro et al. 2020) comprises of a matrix of general linguistic capabilities and test types that facilitate comprehensive test ideation, which leverages a software tool to effectively generate diverse test cases (see also Manerba and Tonelli 2021). Another alternative using existing benchmarks is to use algorithmic audits to identify important demographic and intersectional characteristics and create the necessary test data, for example Buolamwini and Gebru (2018)), who audited commercial gender classifiers by creating an intersectional dataset of gender and skin type, showing that darker-skinned females are the most misclassified group. Complimenting audits, adversarial testing, for example Niven and Kao (2019), and red teaming (Ganguli et al. 2022) can help determine which aspects of the problem space remain challenging and check for the potential harms coming from system biases. In terms of this being a practical guide for research in LLMs, there needs to be some discussion about evaluating the models themselves for ethics, bias and trustworthiness. This is a burgeoning field and we point the reader to a recent taxonomy on LLM evaluation (Chang et al. 2023), in particular Section 3.2.

### 7.2 Harm Evaluation

Despite the ubiquitous nature of the harms of LLMs (Rauh et al. 2022; Weidinger et al. 2021) (further discussed in Section 8.1), the study of such harms has yet to be standardised. Attempts often lack rigour (Blodgett et al. 2020, 2021; Goldfarb-Tarrant et al. 2023). In the following, we briefly present some popular methods for evaluating harms in LLMs, discuss ethical implications and make recommendations. We first categorise approaches by type of harm, then by testing paradigm.

#### 7.2.1 By Harm.

*Safety and Trust.* There are a number of tests designed to evaluate safety in LLMs, which Dinan et al. (2022) provide as a repository to test for safety issues such as the "impostor effect" (where a model gives unsafe advice). Levy et al. (2022) created a benchmark to test for whether a model gives unsafe advice with regards to physical safety. Sun et al. (2023) provide a safety benchmark for Chinese across different safety risks. These safety tests can be useful before deployment to ensure that model safety interventions are successful. However, we caution against being overconfident in their results: a model that performs well in these tests is not necessarily safe in new contexts. Safety behaviour can also be too extreme (for example, failing to respond to "innocent" prompts mentioning "coke" (Röttger et al. 2024)), and Röttger et al. (2024) provide a test suite to identify such exaggerated safety behaviour.

Huang et al. (2023a) detail safety and trustworthiness related issues with LLMs and offer a taxonomy of issues, including factual and reasoning errors, and privacy leaks and data poisoning attacks. Whilst trustworthiness is inherently subjective (Knowles et al. 2023), safety mechanisms e.g. to prevent misinformation are necessary so LLM "behaviour can be assured to be safe and trustable" (Huang et al. 2023a). The authors survey use of Validation and Verification (V&V) tools to assess safety and trustworthiness, and their paper would be a useful reference document for how to integrate V&V tests into LLMs evaluation. Furthermore, Huang et al. (2023b) provide a prompt-based benchmark for trustworthiness which they operationalise as level of toxicity, bias and value alignment, while Tan et al. (2021) propose a six step framework for testing reliability of NLP systems which they relate to safety and fairness. Finally, ALTAI [12] (discussed in Section 3.2) uses questions to prompt evaluation of seven dimensions of trustworthiness, which includes safety.

*Fairness and bias.* Related to the concepts of safety and trust is that of fairness and "bias", typically used (often undefined (Blodgett et al. 2020)) to refer to unwanted differences in a model's representations or outputs as related to different social identities. Unfortunately, bias and fairness testing is hampered by a lack of terminological and methodological precision (Goldfarb-Tarrant et al. 2023; Blodgett et al. 2020, 2021; Cabello, Zee, and Søgaard 2023), so you should carefully define what harm you are trying to measure and why the chosen operationalisation is appropriate (Goldfarb-Tarrant et al. 2023). You will benefit from collaborating with experts from other fields to inform metric design and understand the implications of your findings (Goldfarb-Tarrant et al. 2023; Blodgett et al. 2020). Lack of precision harms validity – that is, whether a metric consistently measures what it claims to measure (Goldfarb-Tarrant et al. 2023; Delobelle et al. 2022; Blodgett et al. 2021); this is evident in the lack of correlation between bias metrics (Delobelle et al. 2022; Dev et al. 2022). You should also be mindful when re-using existing bias metrics in novel contexts, as Goldfarb-Tarrant et al. (2023) find inappropriate re-use of identity lists in new contexts limits validity. Take inspiration from psychology by making careful re-use of metrics as validated by the original creators. This relates to moves to more consistently document models e.g. Model Cards (Mitchell et al. 2019), see Section 6.

Whilst the majority of work on fairness and bias in LLMs has focused on English used in a Western Context (including for multi-lingual models! (Goldfarb-Tarrant et al. 2023)) there is some working looking at fairness in non-US contexts and beyond English (Bhatt et al. 2022; Ramesh, Sitaram, and Choudhury 2023). Evaluating the performance

12 https://ec.europa.eu/newsroom/dae/document.cfm?doc_id=68342

of multi-lingual models introduces a fairness problem in that when choosing a model, performance may differ across languages making it hard to compare between models. A novel solution is provided by Choudhury and Deshpande (2021), who propose selecting the model which maximises the minimum performance.

#### 7.2.2 By Methodology.

*Intrinsic Measures.* A prominent research direction has been to develop intrinsic bias measures which measure proximity of (contextualised) vectors e.g. Caliskan, Bryson, and Narayanan (2017); Tan and Celis (2019); May et al. (2019). Some work has tried to ground such intrinsic measures in social science theory to better align metrics with human judgements (Cao et al. 2022b; Ungless et al. 2022). Intrinsic and extrinsic fairness measures often do not align, suggesting intrinsic metrics have limited validity (Goldfarb-Tarrant et al. 2021; Cao et al. 2022a; Delobelle et al. 2022). Cao et al. (2022a) recommend identifying intrinsic metrics which most closely align with the intended downstream application. There has been research looking to identify intrinsic measures that correlate with extrinsic measures (Orgad, Goldfarb-Tarrant, and Belinkov 2022).

*Prompting (Upstream).* With the rise in generative models, prompting is widely used to detect harms including bias (Goldfarb-Tarrant et al. 2023). For example, Smith et al. (2022) created a vast data set of prompts to test for bias across 13 demographic axes, developed with experts and community members. Some popular prompt data sets have been critiqued for not being grounded in theories of oppression (Blodgett et al. 2021). Typically the output is evaluated using automatic metrics such as sentiment or toxicity score (Goldfarb-Tarrant et al. 2023), but we recommend you also include some human evaluation, which is key to identifying less obvious harms such as "inspiration porn" (content that portrays disabled people as inspiration for abled people) (Gadiraju et al. 2023). Prompting can also be used to probe for e.g. ethical values (Chun and Elkins 2024) and safety (Dinan et al. 2022). Prompting may be used as part of a program of red-teaming, whereby language models are systematically prompted by humans and/or other LLMs, to identify potential sources of harmful output (Perez et al. 2022; Zhuo et al. 2023; Ganguli et al. 2022). Rastogi et al. (2023) propose a tool, AdaTest++, to use LLMs to improve the diversity of prompt tests.

*In-context (Downstream).* Bias and other harms may be measured through performance on downstream tasks and in specific applications. For example there are several data sets designed to test for safety issues in conversational models (Smith et al. 2022; Ung, Xu, and Boureau 2022; Dinan et al. 2022; Barikeri et al. 2021) (which typically rely on prompting). Gupta et al. (2024) propose a cross-task benchmark that tests for bias in LLMs by the performance on a range of tasks, namely QA, sentiment analysis and natural language inference (NLI). Zhang et al. (2023) evaluate the fairness of ChatGPT as a recommender system and propose a new benchmark for this task. Evaluations should be situated to the specific contexts and subjectivities of specific use cases (Blodgett et al. 2020; Talat et al. 2022). Prompts can be crowd-sourced from people who are likely to experience the particular harms that are being evaluated. We therefore reiterate recommendations that you examine the particular ways in which models may cause harms in your use-case, and evaluate them for biases, ideally in equal partnership with people who would be negatively affected by such harms.

*Sociotechnical Safety.* The focus of AI system safety evaluation is typically on the technical components of an AI system, i.e. the individual technical artefacts such as data, model architecture, and sampling. However, it is often the context of this AI system that determines whether a given capability may cause harm. An approach is needed that takes into account human and systemic factors that co-determine risks of harm. Weidinger et al. (2023) define Sociotechnical Safety Evaluation and propose a three layered framework which apart from the Capability Evaluation layer, adds a Human Interaction layer, and a Systemic Impact layer. Sociotechnical safety is already part of the NIST Risk Management Framework.

### 7.3 Key Resources

Do's and Don'ts

- **Do** pair bias metrics that relate to real world (downstream) harms with human evaluation - **don't** rely on quick, quantitative metrics alone, as evaluation in language generation can be unreliable
- **Do** develop and use benchmarks to evaluate concrete, well-scoped and contextualized tasks - **don't** present benchmarks as markers of progress towards general-purpose capabilities
- **Do** carefully reflect on what specific harm you are trying to measure and why the methodology you have created or borrowed is appropriate - **don't** assume bias metrics can be re-used in all new contexts
- **Do** use alternatives to benchmarks which attempt to capture broader capabilities and risks e.g. audits (e.g. Buolamwini and Gebru (2018)), adversarial testing (e.g. Niven and Kao (2019)) and red teaming (Ganguli et al. 2022)
- **Do** involve experts and community members in the evaluation of the models – **don't** rely on your intuitions and initial assumptions alone

Useful Tool(kit)s:

- Tools to facilitate test ideation – Ribeiro et al. (2020)
- Taxonomy of LLM evaluations – Chang et al. (2023) – in particular Section 3.2 on evaluating robustness, ethics, bias, and trustworthiness
- Repository of tests for (English) language generation safety – Dinan et al. (2022)
- Test suite to identifying exaggerated safety behaviour – Röttger et al. (2024)
- Taxonomy of tests for safety and trustworthiness in LLMs – Huang et al. (2023a)
- Framework for testing reliability of NLP systems – Tan et al. (2021)
- Bias tests across hundreds of identities (in English) – Smith et al. (2022)
- Framework for addressing Sociotechnical (contextualised) Safety – Weidinger et al. (2023)

## 8. Deployment

***Overview of ethical issues related to deployment and dissemination of NLP artefacts.***

### 8.1 Likely Harms

***An overview of the likely harms of LLMs.***

With the recent introduction of LLMs and generative AI into society, there has been an increased interest in the application of NLP tools and research.[13] Increases in the application of research, and its outputs, requires a particular attention and care from researchers to avoid enacting harms.

In their broad overview of the harms that arise from generative AI, Solaiman et al. (2024) present seven over-arching categories of social harms from technical systems, including bias, stereotypes, and representational harms (we discuss bias in Section 7.2); cultural values and sensitive content (we discuss best practice for sensitive content in Section 4.2); disparate performances (we discuss fairness and performance in Section 7); environmental costs and carbon emissions (we discuss environmental costs in Section 3.4); privacy and data protections (we discuss consent and safety in Section 4.2); financial costs; and data and content moderation labour (we discuss crowdworkers' rights in Section 5.5). However, in recognition that these cannot be separated from impacts on society, Solaiman et al. (2024) also present categories of "impacts" on society, including trustworthiness and autonomy (which we discuss below), inequality, marginalisation and violence, the concentration of authority, labour and creativity, and ecosystem and environmental impacts. Alongside Solaiman et al. (2024), Weidinger et al. (2021) and Kumar et al. (2022) have also addressed risks of generative AI, finding a subset of the risks and concerns identified by Solaiman et al. (2024). While these sets of authors have focused on generative AI, many of the same concerns—such as bias, stereotypes, and representational harms—have been well documented for other NLP technologies (e.g., Anand et al. 2024; Bolukbasi et al. 2016; Davidson, Bhattacharya, and Weber 2019; De-Arteaga et al. 2019).

A pressing issue is that LLMs suffer from the issue of *hallucination*, or the production of misinformation. Indeed, as Solaiman et al. (2024) remind us, "[increased] access to generating content and potential misinformation, and difficulty distinguishing between human and AI-generated content, poses risks to trust in media and content authenticity," thus presenting a challenge to knowledge-based societies. Training data may then become "polluted" with generated misinformative content which then degrades future performance (Pan et al. 2023). Several researchers have proposed watermarking of LLM-generated content (Kirchenbauer et al. 2023; Grinbaum and Adomaitis 2022). This would help to identify generated content, and also plagiarism (in the sense of using LLMs to write supposedly original content); current detection methods suffer from bias against non-native speakers (Liang et al. 2023). LLMs can also be guilty of plagiarism in the sense of reproducing copyright content. Reproduction of the training data can infringe on both copyright and privacy, as PI is reproduced by the models (Lee et al. 2022; Huang, Shao, and Chang 2022; Lukas et al. 2023) (see Section 5.1 on best practice to avoid this). Here we have given an overview of the likely harms of LLMs, but you

13 As of January 2022, 15% of businesses in the United Kingdom where using some form of AI and another 10% planning to invest in AI (Capital Economics 2022).

should consult Solaiman et al. (2024); Weidinger et al. (2021); Kumar et al. (2022) for more details.

**8.1.1 Dual use.** ***Considering the dual – both negative and positive – use of LLMs.***

The risks of harms arising from dual use of language technologies was first raised by Hovy and Spruit (2016), describing how they can be used for good, e.g., "shed[ing] light on the provenance of historic texts"while also holding potential for negative consequences, e.g., censorship and "endanger[ing] the anonymity of political dissenters." It is often hard to imagine how one's research artefacts might be used for nefarious purposes (though see Section section 3.2 for some toolkits to anticipate harm). Yet, as NLP research is increasingly integrated into consumer- and business-facing products, considerations of how technologies might be misused become increasingly important. Addressing risks of dual use requires in-depth and contextual consideration of the technology and deployment contexts (Kaffee et al. 2023), to which end Kaffee et al. (2023) present a definition of dual use, and a checklist for consideration in research projects. With their checklist, Kaffee et al. (2023) invite practitioners to reflect on the risks of dual use and potential mitigation strategies.

Derczynski et al. (2023) argue for a need for *Risk Cards*, which seek to make clear the risks that can come from models. Risk Cards centre each risk, e.g., the generation of misinformation, rather than the model in which a harm might manifest. They are specifically developed with LLMs in mind, and consider harms that can arise from LLMs, such as representational harms. Risk Cards are best used as living documents, where each new model you develop is subject to tests for a risk, and as the risk is expanded on, e.g. by identifying sub-risks, models are retested. This presents a change from the current practice of only testing models once. Beyond documentation, one can also engage in partial release practices (Kaffee et al. 2023; Solaiman et al. 2019), and using license terms that prohibit specific uses (McDuff et al. 2024; Jernite et al. 2022) (see below).

### 8.2 Release and Beyond

Herein we discuss ethical considerations related to release of a new technology, specifically we discuss ramifications of level of openness; public perceptions of new and even retired technologies, and the impact of new technologies on their context.

**8.2.1 Release Strategy.** Releasing LLMs requires balancing the benefits of different release approaches with the potential risks of misuse. Some LLMs are advertised as being "open" or "open source", though the term "open" lacks an agreed definition and can refer to various aspects of the model release e.g. publicly accessible model parameters ("open weights"), transparent training procedures, open data for training etc (Solaiman et al. 2019; Widder, West, and Whittaker 2023; Groeneveld et al. 2024). Openness enhances transparency, reproducibility and fosters collaborative advancements (Widder, West, and Whittaker 2023; Spirling 2023), allowing for community-driven improvements (Madnani et al. 2017). However, fully open releases may carry significant risks, such as misuse for unintended harmful activities, including after alterations (Arditi et al. 2024). While a model with restricted access can be easily taken down, an openly released model may be impossible to retract. These risks can be mitigated by adopting careful strategies before, during and after deployment. Pre-release audits can help identify biases and security vulnerabilities (Madnani et al. 2017; Ji et al. 2023). The release process should be staged, with an initial release to trusted researchers, followed by a gradual wider release (Solaiman et al. 2019), with post-deployment monitoring to assess new risks (Anderljung et al. 2023). Commercially deployed systems should provide a reporting or complaints procedure.

**8.2.2 Public perception.** Concomitant with product release types, public perception must be considered. One task where this has been explored in detail is automated decision-making (ADM). How fair ADM systems are perceived to be varies between studies (Lee 2018; Araujo 2020), and is influenced by a number of individual factors (Wang, Harper, and Zhu 2020); this is likely true for the fairness of other AI models. Exacerbating these concerns, algorithmic harms can have long lasting consequences on how people understand their lives and AI, even after a technology is "retired" (Ehsan et al. 2022). It is not enough to mitigate harms such as unfairness: the public must also believe these harms to have been mitigated, or their concerns will remain. You must take into account perceptions of your model – not just how it "really works" – when trying to limit its harmful impact. This pertains to release strategy, marketing language, interface design and more.

**8.2.3 Ripples and Coils.** When deploying LLMs, you may fall into common traps well known from previous automated systems, pertaining to fairness, justice, and due process (Selbst et al. 2019). For example the Ripple Effect Trap, when developers fail to consider how technology changes existing social systems. This is particularly relevant to LLMs, as their use can feedback on the data we collect (Pan et al. 2023), the social world we model, and the decisions we make, leading to a looping effect known as data-coiling (Beer 2022). To avoid these traps, design processes should accurately model the specific context for deployment (Selbst et al. 2019), and engage with the conceptual problems and questions that arise from data-coiling (Beer 2022).

### 8.3 Interventions at Inference

***Limited success of interventions at inference time to minimise harm.***

There are a number of ways harmful content from LLMs might be prevented at inference, including guardrails (systems designed to detect when the model might output offensive content, and substitute an alternative), output filtering and prompt modification. Guardrails are widely used on closed-source LLMs such as ChatGPT, Bard etc. Unfortunately, as evidenced by numerous media reports (and on X, formerly Twitter) these guardrails are very brittle (Cuthbertson 2023).

It has been suggested that model output might be improved with regards to harms by simple prompt modification. Natural language interventions using ethical prompts have been unsuccessful for older LLMs (Zhao et al. 2021). They have also been explored for text-to-image models (Bansal et al. 2022), with mixed success (Shin et al. 2024). Prompt modification may also be happening "behind the scenes" as part of the guard rails discussed above, as has been demonstrated for Microsoft Bing (Edwards 2023). However, prompt modification can also be problematic if done insensitively. Ungless, Ross, and Lauscher (2023) find transgender people reject the removal of sensitive terms to prevent harm, and Vynck and Tiku (2024) document significant push back to Google's decision to modify prompts to increase ethnic and gender diversity including in inappropriate contexts.

Finally, a very simple attempt to lessen harm at inference is to warn of the potential for offensive or inaccurate output.[14] However research suggests warning messages are unpopular amongst marginalised groups (at least for text-to-image models (Ungless, Ross, and Lauscher 2023)). Further, whilst research suggests warning labels may be useful for possible AI-generated misinformation (Wittenberg et al. 2024), this is currently no evidence to support that these warnings actually prevent harm of offensive content. They are far too vague to enable people to avoid being exposed to particular upsetting content (e.g. offensive content could be racist stereotypes, graphic gore, sexually explicit etc).

It seems that guardrails and other interventions at inference are brittle and have shown limited success at addressing ethical issues "baked into" LLMs.

### 8.4 Using LLMs to replace humans

***Ethical ramifications of using LLM to replace humans at various different tasks.***

The proliferation of LLMs makes it tempting to replace humans by LLMs to increase efficiency. LLMs are being considered even for professions with high levels of impact, such as content creation, journalism and creative writing (WGA Negotiating Comittee 2023), as well as education (Walczak and Cellary 2023; Kasneci et al. 2023; Sok and Heng 2023). While automating some repetitive aspect of a person's job might have ethical justification, replacing humans with machines could lead to significant problems with unemployment and could do fundamental damage to society. Journalists and educators do far more than deliver information efficiently and serious study of stakeholders and risks should be undertaken before any project is approved.

That said in education, LLMs show some promise for automated creation of educational content, improvement of student engagement and personalisation of the learning experience (Kasneci et al. 2023). However, responsible application of these tools requires digital literacy as well as understanding of capabilities and limitations by both teachers and students (Walczak and Cellary 2023; Kasneci et al. 2023). Deployment in this context demands adherence to privacy, security and regulatory requirements (Kasneci et al. 2023) while universities should put more emphasis on teaching life-long self learning skills (Walczak and Cellary 2023).

LLMs have also been considered for replacing human participants in domains such as psychology, user research and AI technology development. This holds the promise of cutting costs, avoiding potential participant harms and increasing (apparent) demographic diversity (Agnew et al. 2024). Notably, Chiang and Lee (2023) reported success with replacing humans for evaluation; they find LLM story evaluation was consistent with evaluations by human experts. However, this result is contrasted by a range of other studies reporting major shortcomings. Aher, Arriaga, and Kalai (2022) find that some LLMs have a tendency to give unhumanly accurate answers. Similarly, Bavaresco et al. (2024) observe shortcomings in replicating human judgements, where correlations between human and LLM judgements varied considerably across data sets. LLMs also performed poorly when evaluated on the labelling of a fairness benchmark from a community survey (Felkner, Thompson, and May 2024). These studies question the readiness of LLMs to replace human participants. Going further, Agnew et al. (2024) argue that LLM replacement of participants undermines foundational values of work

14 e.g. ChatGPT has a warning message "ChatGPT can make mistakes. Check important info."

with human participants, namely representation of participants' interests, inclusion of participants in the development process and understanding. These shortcomings are fundamental and cannot be fixed with better training procedures (Agnew et al. 2024).

An alternative avenue to explore the potential of LLMs is to keep humans in the loop. Shneiderman (2020) propose a framework called Human-Centered Artificial Intelligence, which aims to use AI techniques in the right way to increase human performance and ensure that AI systems are reliable, safe and trustworthy. The framework also defines scenarios where full control or automation is justified.

### 8.5 Dissemination

Your responsibility for ethical outcomes does not end when you finish work on a given system (Widder and Nafus 2023). In most cases our intention is that others will further develop what we have done. We share responsibility for that ongoing work, in part because the way in which we distribute our work influences how others are predisposed to interact with it. The most obvious responsibility is in making sure future practitioners have the information they require to make informed ethical decisions. This highlights the importance of openness about processes, data (Mieskes 2017), and other resources used (Schwartz et al. 2019). Toolkits like Model Cards and Datasheets (Mitchell et al. 2019; Gebru et al. 2021) are a good start, but we shouldn't let our responsible practice deteriorate to just another technical documentation task (Widder 2024).

Both in descriptions of our own work, and in discussion of the field as a whole, the way we talk about our technologies can have a wider impact on understanding. While the demands of competitive funding calls may lead us to oversell, chains of this happening over and over again, catalysed by the media, contribute to hype cycles (Markelius et al. 2024). Ultimately this means decision makers in government and in wider society are left misinformed about LLM capabilities, leading to harmful decisions about their governance or deployment.

Finally, it is worth remembering that you can be a role model: how you demonstrate the centrality of ethical thinking in your work will in turn encourage others around you to think differently about it. This includes your own use of LLMs (Guleria et al. 2023; Lund et al. 2023).

### 8.6 Key Resources

Do's and Don'ts

- **Do** consider integrating watermarking into your generative models – **don't** rely on supervised detection models alone
- **Do** pre-release audits to identify biases and security vulnerabilities (Madnani et al. 2017) – **don't** put the onus on marginalised people to discover harms
- **Do** release LLMs in stages, with an initial release to trusted researchers, followed by a gradual wider release (Solaiman et al. 2019) – **don't** forget the model will change its own environment in terms of both training data and people's expectations

- **Do** continually monitor post-deployment to assess new risks and create a complaints procedure (Anderljung et al. 2023) – **don't** count on brittle guardrails to prevent harm
- **Do** consider how AI might enhance human experience of work, as well as performance – **don't** assume LLMs can effectively replace human participants
- **Do** consider how the public perceive your technology – **don't** contribute to the hype cycle

Useful Tool(kit)s:

- Framework to encourage AI that enhances rather than replaces human performance – Shneiderman (2020)
- Overview of harms and ramifications of generative AI technologies – Solaiman et al. (2024)
- A definition of dual use, and a checklist for consideration in research projects – Kaffee et al. (2023)
- Documentation methodology for risks of LLMs, that could be adapted to document dual use impact – Derczynski et al. (2023)

## 9. Conclusion

Whilst there is already a significant body of work looking at the ethical considerations of large language models (LLMs), to date there has been no work that integrates these resources into a single guide; in this Whitepaper, we attempted this ambitious goal. Whilst by no means complete in its coverage of ethical issues related to LLMs, we hope our clear Do's and Don'ts and selection of tool kits has been helpful in translating the ethics literature into concrete recommendations for computer scientists. We present this as an open resource for those conducting research, or those tasked with evaluating the ethical implications of others' works.

## Acknowledgements

This work was improved with insightful feedback from Shannon Vallor and Jacquie Rowe. This work was partly funded by the Generative AI Laboratory (GAIL), University of Edinburgh. Alexandra Birch was partly funded by the UK Research and Innovation (UKRI) under the UK government's Horizon Europe funding guarantee [grant number 10039436 (Utter)]. Eddie L. Ungless was supported by the UKRI Centre for Doctoral Training in Natural Language Processing, funded by the UKRI (grant EP/S022481/1) and the University of Edinburgh, School of Informatics.

## References

Abercrombie, Gavin, Amanda Cercas Curry, Tanvi Dinkar, Verena Rieser, and Zeerak Talat. 2023. Mirages. on anthropomorphism in dialogue systems. In *Proceedings of the 2023 Conference on Empirical Methods in Natural Language Processing*, pages 4776–4790, Association for Computational Linguistics, Singapore.

Abercrombie, Gavin, Nikolas Vitsakis, Aiqi Jiang, and Ioannis Konstas. 2024. Revisiting annotation of online gender-based violence. In *Proceedings of the 3rd Workshop on Perspectivist*

*Approaches to NLP (NLPerspectives) @ LREC-COLING 2024*, pages 31–41, ELRA and ICCL, Torino, Italia.

Agnew, William, A. Stevie Bergman, Jennifer Chien, Mark Díaz, Seliem El-Sayed, Jaylen Pittman, Shakir Mohamed, and Kevin R. McKee. 2024. The illusion of artificial inclusion. ArXiv:2401.08572 [cs].

Aher, Gati, RosaI Arriaga, and Adam Tauman Kalai. 2022. Using Large Language Models to Simulate Multiple Humans. *ArXiv*, abs/2208.10264.

Al Kuwatly, Hala, Maximilian Wich, and Georg Groh. 2020. Identifying and Measuring Annotator Bias Based on Annotators' Demographic Characteristics. In *Proceedings of the Fourth Workshop on Online Abuse and Harms*, pages 184–190, Association for Computational Linguistics, Online.

Allahbakhsh, Mohammad, Boualem Benatallah, Aleksandar Ignjatovic, Hamid Reza Motahari-Nezhad, Elisa Bertino, and Schahram Dustdar. 2013. Quality control in crowdsourcing systems: Issues and directions. *IEEE Internet Computing*, 17(2):76–81.

Anand, Abhishek, Negar Mokhberian, Prathyusha Naresh Kumar, Anweasha Saha, Zihao He, Ashwin Rao, Fred Morstatter, and Kristina Lerman. 2024. Don't Blame the Data, Blame the Model: Understanding Noise and Bias When Learning from Subjective Annotations.

Anderljung, Markus, Joslyn Barnhart, Jade Leung, Anton Korinek, Cullen O'Keefe, Jess Whittlestone, Shahar Avin, Miles Brundage, Justin Bullock, Duncan Cass-Beggs, et al. 2023. Frontier ai regulation: Managing emerging risks to public safety. *arXiv preprint arXiv:2307.03718*.

Anwar, Usman, Abulhair Saparov, Javier Rando, Daniel Paleka, Miles Turpin, Peter Hase, Ekdeep Singh Lubana, Erik Jenner, Stephen Casper, Oliver Sourbut, et al. 2024. Foundational challenges in assuring alignment and safety of large language models. *arXiv preprint arXiv:2404.09932*.

Araujo, Theo. 2020. In AI we trust? Perceptions about automated decision-making by artificial intelligence. *AI & SOCIETY*, (35):13.

Arditi, Andy, O Balcells, A Syed, W Gurnee, and N Nanda. 2024. Refusal in llms is mediated by a single direction. In *Alignment Forum*, page 15.

Askell, Amanda, Yuntao Bai, Anna Chen, Dawn Drain, Deep Ganguli, Tom Henighan, Andy Jones, Nicholas Joseph, Ben Mann, Nova DasSarma, et al. 2021. A general language assistant as a laboratory for alignment. *arXiv preprint arXiv:2112.00861*.

Bannour, Nesrine, Sahar Ghannay, Aurélie Névéol, and Anne-Laure Ligozat. 2021. Evaluating the carbon footprint of NLP methods: a survey and analysis of existing tools. In *Proceedings of the Second Workshop on Simple and Efficient Natural Language Processing*, pages 11–21, Association for Computational Linguistics, Virtual.

Bansal, Hritik, Da Yin, Masoud Monajatipoor, and Kai-Wei Chang. 2022. How well can Text-to-Image Generative Models understand Ethical Natural Language Interventions? In *Proceedings of the 2022 Conference on Empirical Methods in Natural Language Processing*, pages 1358–1370, Association for Computational Linguistics, Abu Dhabi, United Arab Emirates.

Barikeri, Soumya, Anne Lauscher, Ivan Vulić, and Goran Glavaš. 2021. RedditBias: A Real-World Resource for Bias Evaluation and Debiasing of Conversational Language Models. In *Proceedings of the 59th Annual Meeting of the Association for Computational Linguistics and the 11th International Joint Conference on Natural Language Processing (Volume 1: Long Papers)*, pages 1941–1955, Association for Computational Linguistics, Online.

Basile, Valerio. 2020. It's the end of the gold standard as we know it: Leveraging non-aggregated data for better evaluation and explanation of subjective tasks. In *International Conference of the Italian Association for Artificial Intelligence*, pages 441–453, Springer.

Bavaresco, Anna, Raffaella Bernardi, Leonardo Bertolazzi, Desmond Elliott, Raquel Fernández, Albert Gatt, Esam Ghaleb, Mario Giulianelli, Michael Hanna, Alexander Koller, André F. T. Martins, Philipp Mondorf, Vera Neplenbroek, Sandro Pezzelle, Barbara Plank, David Schlangen, Alessandro Suglia, Aditya K. Surikuchi, Ece Takmaz, and Alberto Testoni. 2024. LLMs instead of Human Judges? A Large Scale Empirical Study across 20 NLP Evaluation Tasks. ArXiv:2406.18403 [cs].

Bederson, Benjamin B. and Alexander J. Quinn. 2011. Web workers unite! addressing challenges of online laborers. *CHI '11 Extended Abstracts on Human Factors in Computing Systems*, pages 97–106. Conference Name: CHI '11: CHI Conference on Human Factors in Computing Systems ISBN: 9781450302685 Place: Vancouver BC Canada Publisher: ACM.

Beer, David. 2022. The problem of researching a recursive society: Algorithms, data coils and the looping of the social. *Big Data & Society*, 9(2):20539517221104997. Publisher: SAGE Publications Ltd.

Bender, Emily M. and Batya Friedman. 2018. Data Statements for Natural Language Processing: Toward Mitigating System Bias and Enabling Better Science. *Transactions of the Association for Computational Linguistics*, 6:587–604.

Bender, Emily M., Timnit Gebru, Angelina McMillan-Major, and Shmargaret Shmitchell. 2021. On the Dangers of Stochastic Parrots: Can Language Models Be Too Big? . In *Proceedings of the 2021 ACM Conference on Fairness, Accountability, and Transparency*, pages 610–623, ACM, Virtual Event Canada.

Benjamin, Garfield. 2021. What we do with data: a performative critique of data 'collection'. *Internet Policy Review*, 10(4).

Benotti, Luciana, Karën Fort, Min-Yen Kan, and Yulia Tsvetkov. 2023. Understanding Ethics in NLP Authoring and Reviewing. In *Proceedings of the 17th Conference of the European Chapter of the Association for Computational Linguistics: Tutorial Abstracts*, pages 19–24, Association for Computational Linguistics, Dubrovnik, Croatia.

Benton, Adrian, Glen Coppersmith, and Mark Dredze. 2017. Ethical Research Protocols for Social Media Health Research. In *Proceedings of the First ACL Workshop on Ethics in Natural Language Processing*, pages 94–102, Association for Computational Linguistics, Valencia, Spain.

Bhatt, Shaily, Sunipa Dev, Partha Talukdar, Shachi Dave, and Vinodkumar Prabhakaran. 2022. Re-contextualizing Fairness in NLP: The Case of India. ArXiv:2209.12226 [cs].

Biester, Laura, Vanita Sharma, Ashkan Kazemi, Naihao Deng, Steven Wilson, and Rada Mihalcea. 2022. Analyzing the effects of annotator gender across NLP tasks. In *Proceedings of the 1st Workshop on Perspectivist Approaches to NLP @LREC2022*, pages 10–19, European Language Resources Association, Marseille, France.

Bird, Charlotte, Eddie Ungless, and Atoosa Kasirzadeh. 2023. Typology of Risks of Generative Text-to-Image Models. In *Proceedings of the 2023 AAAI/ACM Conference on AI, Ethics, and Society*, AIES '23, pages 396–410, Association for Computing Machinery, New York, NY, USA.

Bird, Steven. 2020. Decolonising Speech and Language Technology. In *Proceedings of the 28th International Conference on Computational Linguistics*, pages 3504–3519, International Committee on Computational Linguistics, Barcelona, Spain (Online).

Birhane, Abeba, Pratyusha Kalluri, Dallas Card, William Agnew, Ravit Dotan, and Michelle Bao. 2022. The Values Encoded in Machine Learning Research. In *2022 ACM Conference on Fairness, Accountability, and Transparency*, pages 173–184, ACM, Seoul Republic of Korea.

Blodgett, Su Lin, Solon Barocas, Hal Daumé III, and Hanna Wallach. 2020. Language (Technology) is Power: A Critical Survey of "Bias" in NLP. In *Proceedings of the 58th Annual Meeting of the Association for Computational Linguistics*, pages 5454–5476, Association for Computational Linguistics, Online.

Blodgett, Su Lin, Gilsinia Lopez, Alexandra Olteanu, Robert Sim, and Hanna Wallach. 2021. Stereotyping Norwegian Salmon: An Inventory of Pitfalls in Fairness Benchmark Datasets.

Blodgett, Su Lin and Brendan O'Connor. 2017. Racial Disparity in Natural Language Processing: A Case Study of Social Media African-American English. *arXiv:1707.00061 [cs]*. ArXiv: 1707.00061.

Bolukbasi, Tolga, Kai-Wei Chang, James Y Zou, Venkatesh Saligrama, and Adam T Kalai. 2016. Man is to Computer Programmer as Woman is to Homemaker? Debiasing Word Embeddings. In *Advances in Neural Information Processing Systems*, volume 29, Curran Associates, Inc.

Buolamwini, Joy and Timnit Gebru. 2018. Gender Shades: Intersectional Accuracy Disparities in Commercial Gender Classification. In *Conference on Fairness, Accountability and Transparency*, pages 77–91, PMLR. ISSN: 2640-3498.

Buçinca, Zana, Chau Minh Pham, Maurice Jakesch, Marco Tulio Ribeiro, Alexandra Olteanu, and Saleema Amershi. 2023. AHA!: Facilitating AI Impact Assessment by Generating Examples of Harms. *ArXiv*, abs/2306.03280.

Cabello, Laura, Anna Katrine van Zee, and Anders Søgaard. 2023. On the Independence of Association Bias and Empirical Fairness in Language Models. *Proceedings of the 2023 ACM Conference on Fairness, Accountability, and Transparency*.

Caglayan, Ozan, Pranava Madhyastha, and Lucia Specia. 2020. Curious Case of Language Generation Evaluation Metrics: A Cautionary Tale. In *Proceedings of the 28th International Conference on Computational Linguistics*, pages 2322–2328, International Committee on Computational Linguistics, Barcelona, Spain (Online).

Calabrese, Agostina, Michele Bevilacqua, Björn Ross, Rocco Tripodi, and Roberto Navigli. 2021. AAA: Fair Evaluation for Abuse Detection Systems Wanted. In *13th ACM Web Science Conference 2021*, pages 243–252, ACM, Virtual Event United Kingdom.

Caliskan, Aylin, Joanna J. Bryson, and Arvind Narayanan. 2017. Semantics derived automatically from language corpora contain human-like biases. *Science*, 356(6334):183–186. Publisher: American Association for the Advancement of Science Section: Reports.

Cao, Yang, Yada Pruksachatkun, Kai-Wei Chang, Rahul Gupta, Varun Kumar, Jwala Dhamala, and Aram Galstyan. 2022a. On the Intrinsic and Extrinsic Fairness Evaluation Metrics for Contextualized Language Representations. In *Proceedings of the 60th Annual Meeting of the Association for Computational Linguistics (Volume 2: Short Papers)*, pages 561–570, Association for Computational Linguistics, Dublin, Ireland.

Cao, Yang Trista, Anna Sotnikova, Hal Daum'e, Rachel Rudinger, and Linda X. Zou. 2022b. Theory-Grounded Measurement of U.S. Social Stereotypes in English Language Models. In *North American Chapter of the Association for Computational Linguistics*.

Capital Economics. 2022. AI activity in UK businesses: Executive Summary. Commissioned Report, Department for Digital, Culture, Media & Sport, London, UK.

Carlini, Nicholas, Florian Tramer, Eric Wallace, Matthew Jagielski, Ariel Herbert-Voss, Katherine Lee, Adam Roberts, Tom Brown, Dawn Song, Ulfar Erlingsson, Alina Oprea, and Colin Raffel. 2021. Extracting Training Data from Large Language Models. ArXiv:2012.07805 [cs].

Caselli, Tommaso, Roberto Cibin, Costanza Conforti, Enrique Encinas, and Maurizio Teli. 2021. Guiding Principles for Participatory Design-inspired Natural Language Processing. In *Proceedings of the 1st Workshop on NLP for Positive Impact*, pages 27–35, Association for Computational Linguistics, Online.

Chang, Yu-Chu, Xu Wang, Jindong Wang, Yuanyi Wu, Kaijie Zhu, Hao Chen, Linyi Yang, Xiaoyuan Yi, Cunxiang Wang, Yidong Wang, Weirong Ye, Yue Zhang, Yi Chang, Philip S. Yu, Qian Yang, and Xingxu Xie. 2023. A Survey on Evaluation of Large Language Models. *ArXiv*, abs/2307.03109.

Chiang, Cheng-Han and Hung-yi Lee. 2023. Can Large Language Models Be an Alternative to Human Evaluations? In *Annual Meeting of the Association for Computational Linguistics*.

Chmielewski, Michael and Sarah C. Kucker. 2020. An mturk crisis? shifts in data quality and the impact on study results. *Social Psychological and Personality Science*, 11(4):464–473.

Choudhury, Monojit and Amit Deshpande. 2021. How Linguistically Fair Are Multilingual Pre-Trained Language Models? In *AAAI Conference on Artificial Intelligence*.

Chun, Jon and Katherine Elkins. 2024. Informed AI Regulation: Comparing the Ethical Frameworks of Leading LLM Chatbots Using an Ethics-Based Audit to Assess Moral Reasoning and Normative Values. Publisher: [object Object] Version Number: 1.

Courty, Benoit, Victor Schmidt, Sasha Luccioni, Goyal-Kamal, MarionCoutarel, Boris Feld, Jérémy Lecourt, LiamConnell, Amine Saboni, Inimaz, supatomic, Mathilde Léval, Luis Blanche, Alexis Cruveiller, ouminasara, Franklin Zhao, Aditya Joshi, Alexis Bogroff, Hugues de Lavoreille, Niko Laskaris, Edoardo Abati, Douglas Blank, Ziyao Wang, Armin Catovic, Marc Alencon, Michał Stęchły, Christian Bauer, Lucas Otávio N. de Araújo, JPW, and MinervaBooks. 2024. mlco2/codecarbon: v2.4.1.

Cuthbertson, Anthony. 2023. ChatGPT "grandma exploit" helps people pirate software. Publication Title: The Independent.

Daniel, Florian, Pavel Kucherbaev, Cinzia Cappiello, Boualem Benatallah, and Mohammad Allahbakhsh. 2018. Quality control in crowdsourcing: A survey of quality attributes, assessment techniques, and assurance actions. *ACM Comput. Surv.*, 51(1).

Danks, David. 2022. Digital ethics as translational ethics. In *Applied ethics in a digital world*. IGI Global, pages 1–15.

Davani, Aida, Mark Díaz, Dylan Baker, and Vinodkumar Prabhakaran. 2024. Disentangling perceptions of offensiveness: Cultural and moral correlates. In *Proceedings of the 2024 ACM Conference on Fairness, Accountability, and Transparency*, FAccT '24, page 2007–2021, Association for Computing Machinery, New York, NY, USA.

Davani, Aida Mostafazadeh, Mohammad Atari, Brendan Kennedy, and Morteza Dehghani. 2023. Hate speech classifiers learn normative social stereotypes. *Transactions of the Association for Computational Linguistics*, 11:300–319.

Davidson, Thomas, Debasmita Bhattacharya, and Ingmar Weber. 2019. Racial Bias in Hate Speech and Abusive Language Detection Datasets. In *Proceedings of the Third Workshop on Abusive Language Online*, pages 25–35, Association for Computational Linguistics, Florence,

Italy.
De-Arteaga, Maria, Alexey Romanov, Hanna Wallach, Jennifer Chayes, Christian Borgs, Alexandra Chouldechova, Sahin Geyik, Krishnaram Kenthapadi, and Adam Tauman Kalai. 2019. Bias in Bios: A Case Study of Semantic Representation Bias in a High-Stakes Setting. In *Proceedings of the Conference on Fairness, Accountability, and Transparency*, pages 120–128, ACM, Atlanta GA USA.
Delobelle, Pieter, Ewoenam Kwaku Tokpo, Toon Calders, and Bettina Berendt. 2022. Measuring fairness with biased rulers: A comparative study on bias metrics for pre-trained language models. In *Proceedings of the 2022 Conference of the North American Chapter of the Association for Computational Linguistics*, pages 1693–1706, Association for Computational Linguistics.
Derczynski, Leon, Hannah Rose Kirk, Vidhisha Balachandran, Sachin Kumar, Yulia Tsvetkov, M. R. Leiser, and Saif Mohammad. 2023. Assessing Language Model Deployment with Risk Cards. Publisher: [object Object] Version Number: 1.
Dev, Sunipa, Jaya Goyal, Dinesh Tewari, Shachi Dave, and Vinodkumar Prabhakaran. 2024. Building socio-culturally inclusive stereotype resources with community engagement. *Advances in Neural Information Processing Systems*, 36.
Dev, Sunipa, Masoud Monajatipoor, Anaelia Ovalle, Arjun Subramonian, Jeff Phillips, and Kai-Wei Chang. 2021. Harms of Gender Exclusivity and Challenges in Non-Binary Representation in Language Technologies. In *Proceedings of the 2021 Conference on Empirical Methods in Natural Language Processing*, pages 1968–1994, Association for Computational Linguistics, Online and Punta Cana, Dominican Republic.
Dev, Sunipa, Emily Sheng, Jieyu Zhao, Aubrie Amstutz, Jiao Sun, Yu Hou, Mattie Sanseverino, Jiin Kim, Akihiro Nishi, Nanyun Peng, and Kai-Wei Chang. 2022. On Measures of Biases and Harms in NLP. In *Findings of the Association for Computational Linguistics: AACL-IJCNLP 2022*, pages 246–267, Association for Computational Linguistics, Online only.
Devinney, Hannah, Jenny Björklund, and Henrik Björklund. 2022. Theories of "Gender" in NLP Bias Research. In *2022 ACM Conference on Fairness, Accountability, and Transparency*, FAccT '22, pages 2083–2102, Association for Computing Machinery, New York, NY, USA.
Dinan, Emily, Gavin Abercrombie, A. Bergman, Shannon Spruit, Dirk Hovy, Y-Lan Boureau, and Verena Rieser. 2022. SafetyKit: First Aid for Measuring Safety in Open-domain Conversational Systems. In *Proceedings of the 60th Annual Meeting of the Association for Computational Linguistics (Volume 1: Long Papers)*, pages 4113–4133, Association for Computational Linguistics, Dublin, Ireland.
Dodge, Jesse, Taylor Prewitt, Remi Tachet Des Combes, Erika Odmark, Roy Schwartz, Emma Strubell, Alexandra Sasha Luccioni, Noah A. Smith, Nicole DeCario, and Will Buchanan. 2022. Measuring the Carbon Intensity of AI in Cloud Instances. In *2022 ACM Conference on Fairness, Accountability, and Transparency*, pages 1877–1894, ACM, Seoul Republic of Korea.
Dunn, Jonathan. 2020. Mapping languages: The Corpus of Global Language Use. *Language Resources and Evaluation*, 54(4):999–1018.
Edwards, Benj. 2023. AI-powered Bing Chat spills its secrets via prompt injection attack [Updated]. Publication Title: Ars Technica.
Ehsan, Upol, Ranjit Singh, Jacob Metcalf, and Mark Riedl. 2022. The Algorithmic Imprint. In *Proceedings of the 2022 ACM Conference on Fairness, Accountability, and Transparency*, FAccT '22, pages 1305–1317, Association for Computing Machinery, New York, NY, USA.
Elazar, Yanai, Akshita Bhagia, Ian Magnusson, Abhilasha Ravichander, Dustin Schwenk, Alane Suhr, Pete Walsh, Dirk Groeneveld, Luca Soldaini, Sameer Singh, Hanna Hajishirzi, Noah A. Smith, and Jesse Dodge. 2024. What's In My Big Data? ArXiv:2310.20707 [cs].
ElSherief, Mai, Caleb Ziems, David Muchlinski, Vaishnavi Anupindi, Jordyn Seybolt, Munmun De Choudhury, and Diyi Yang. 2021. Latent Hatred: A Benchmark for Understanding Implicit Hate Speech. In *Proceedings of the 2021 Conference on Empirical Methods in Natural Language Processing*, pages 345–363, Association for Computational Linguistics, Online and Punta Cana, Dominican Republic.
Excell, Elizabeth and Noura Al Moubayed. 2021. Towards Equal Gender Representation in the Annotations of Toxic Language Detection. In *Proceedings of the 3rd Workshop on Gender Bias in Natural Language Processing*, pages 55–65, Association for Computational Linguistics, Online.
Eyal, Peer, Rothschild David, Gordon Andrew, Evernden Zak, and Damer Ekaterina. 2021. Data quality of platforms and panels for online behavioral research. *Behavior research methods*, pages 1–20.

Felkner, Virginia K., Jennifer A. Thompson, and Jonathan May. 2024. GPT is Not an Annotator: The Necessity of Human Annotation in Fairness Benchmark Construction. ArXiv:2405.15760 [cs].

Feng, Shangbin, Chan Young Park, Yuhan Liu, and Yulia Tsvetkov. 2023. From pretraining data to language models to downstream tasks: Tracking the trails of political biases leading to unfair NLP models. In *Proceedings of the 61st Annual Meeting of the Association for Computational Linguistics (Volume 1: Long Papers)*, pages 11737–11762, Association for Computational Linguistics, Toronto, Canada.

Fiesler, Casey, Michael Zimmer, Nicholas Proferes, Sarah Gilbert, and Naiyan Jones. 2024. Remember the Human: A Systematic Review of Ethical Considerations in Reddit Research. *Proc. ACM Hum.-Comput. Interact.*, 8(GROUP):5:1–5:33.

Fleisig, Eve, Rediet Abebe, and Dan Klein. 2023. When the majority is wrong: Modeling annotator disagreement for subjective tasks. In *Proceedings of the 2023 Conference on Empirical Methods in Natural Language Processing*, pages 6715–6726, Association for Computational Linguistics, Singapore.

Fort, Karën, Gilles Adda, and K. Bretonnel Cohen. 2011. Last Words: Amazon Mechanical Turk: Gold Mine or Coal Mine? *Computational Linguistics*, 37(2):413–420. Place: Cambridge, MA Publisher: MIT Press.

Fort, Karën, Gilles Adda, Benoît Sagot, Joseph Mariani, and Alain Couillault. 2014. Crowdsourcing for Language Resource Development: Criticisms About Amazon Mechanical Turk Overpowering Use. In *Human Language Technology Challenges for Computer Science and Linguistics*, pages 303–314, Springer International Publishing, Cham.

Fortuna, Paula, Monica Dominguez, Leo Wanner, and Zeerak Talat. 2022. Directions for NLP Practices Applied to Online Hate Speech Detection. In *Proceedings of the 2022 Conference on Empirical Methods in Natural Language Processing*, pages 11794–11805, Association for Computational Linguistics, Abu Dhabi, United Arab Emirates.

Fukuda-Parr, Sakiko and Elizabeth Gibbons. 2021. Emerging consensus on 'ethical AI': Human rights critique of stakeholder guidelines. *Global Policy*, 12:32–44. Publisher: Wiley Online Library.

Fussell, Sidney. 2019. How an Attempt at Correcting Bias in Tech Goes Wrong. Section: Technology.

Gabriel, Iason. 2020. Artificial intelligence, values, and alignment. *Minds and machines*, 30(3):411–437.

Gadiraju, Vinitha, Shaun Kane, Sunipa Dev, Alex Taylor, Ding Wang, Emily Denton, and Robin Brewer. 2023. "I wouldn't say offensive but...": Disability-Centered Perspectives on Large Language Models. In *Proceedings of the 2023 ACM Conference on Fairness, Accountability, and Transparency*, FAccT '23, pages 205–216, Association for Computing Machinery, New York, NY, USA.

Gallegos, Isabel O, Ryan A Rossi, Joe Barrow, Md Mehrab Tanjim, Sungchul Kim, Franck Dernoncourt, Tong Yu, Ruiyi Zhang, and Nesreen K Ahmed. 2024. Bias and fairness in large language models: A survey. *Computational Linguistics*, pages 1–79.

Ganguli, Deep, Amanda Askell, Nicholas Schiefer, Thomas Liao, Kamil e Lukovsiut.e, Anna Chen, Anna Goldie, Azalia Mirhoseini, Catherine Olsson, Danny Hernandez, Dawn Drain, Dustin Li, Eli Tran-Johnson, Ethan Perez, John Kernion, Jamie Kerr, Jared Mueller, Joshua D. Landau, Kamal Ndousse, Karina Nguyen, Liane Lovitt, Michael Sellitto, Nelson Elhage, Noem'i Mercado, Nova DasSarma, Robert Lasenby, Robin Larson, Sam Ringer, Sandipan Kundu, Saurav Kadavath, Scott Johnston, Shauna Kravec, Sheer El Showk, Tamera Lanham, Timothy Telleen-Lawton, Tom Henighan, Tristan Hume, Yuntao Bai, Zac Hatfield-Dodds, Benjamin Mann, Dario Amodei, Nicholas Joseph, Sam McCandlish, Tom B. Brown, Christopher Olah, Jack Clark, Sam Bowman, and Jared Kaplan. 2023. The Capacity for Moral Self-Correction in Large Language Models. *ArXiv*, abs/2302.07459.

Ganguli, Deep, Liane Lovitt, John Kernion, Amanda Askell, Yuntao Bai, Saurav Kadavath, Benjamin Mann, Ethan Perez, Nicholas Schiefer, Kamal Ndousse, Andy Jones, Sam Bowman, Anna Chen, Tom Conerly, Nova DasSarma, Dawn Drain, Nelson Elhage, Sheer El-Showk, Stanislav Fort, Zachary Dodds, Tom Henighan, Danny Hernandez, Tristan Hume, Josh Jacobson, Scott Johnston, Shauna Kravec, Catherine Olsson, Sam Ringer, Eli Tran-Johnson, Dario Amodei, Tom B. Brown, Nicholas Joseph, Sam McCandlish, Christopher Olah, Jared Kaplan, and Jack Clark. 2022. Red Teaming Language Models to Reduce Harms: Methods, Scaling Behaviors, and Lessons Learned. *ArXiv*, abs/2209.07858.

Gebru, Timnit, Jamie Morgenstern, Briana Vecchione, Jennifer Wortman Vaughan, Hanna Wallach, Hal Daumé III, and Kate Crawford. 2020. Datasheets for Datasets. *arXiv:1803.09010 [cs]*.

Gebru, Timnit, Jamie Morgenstern, Briana Vecchione, Jennifer Wortman Vaughan, Hanna Wallach, Hal Daumé Iii, and Kate Crawford. 2021. Datasheets for datasets. *Communications of the ACM*, 64(12):86–92.

Gitelman, Lisa and Virginia Jackson. 2013. Introduction. In Lisa Gitelman, editor, *Raw Data Is an Oxymoron*. MIT Press, pages 1–14.

Gold, Zachary and Mark Latonero. 2017. Robots welcome: Ethical and legal considerations for web crawling and scraping. *Washington Journal of Law, Technology & Arts*, 13:275.

Goldfarb-Tarrant, Seraphina, Rebecca Marchant, Ricardo Muñoz Sánchez, Mugdha Pandya, and Adam Lopez. 2021. Intrinsic Bias Metrics Do Not Correlate with Application Bias. In *Proceedings of the 59th Annual Meeting of the Association for Computational Linguistics and the 11th International Joint Conference on Natural Language Processing (Volume 1: Long Papers)*, pages 1926–1940, Association for Computational Linguistics, Online.

Goldfarb-Tarrant, Seraphina, Eddie Ungless, Esma Balkir, and Su Lin Blodgett. 2023. This prompt is measuring <mask>: evaluating bias evaluation in language models. In *Findings of the Association for Computational Linguistics: ACL 2023*, pages 2209–2225, Association for Computational Linguistics, Toronto, Canada.

Gonen, Hila and Yoav Goldberg. 2019. Lipstick on a Pig: Debiasing Methods Cover up Systematic Gender Biases in Word Embeddings But do not Remove Them. In *Proceedings of the 2019 Conference of the North American Chapter of the Association for Computational Linguistics: Human Language Technologies, Volume 1 (Long and Short Papers)*, pages 609–614, Association for Computational Linguistics, Minneapolis, Minnesota.

Grinbaum, Alexei and Laurynas Adomaitis. 2022. The Ethical Need for Watermarks in Machine-Generated Language. *ArXiv*, abs/2209.03118.

Groeneveld, Dirk, Iz Beltagy, Pete Walsh, Akshita Bhagia, Rodney Kinney, Oyvind Tafjord, Ananya Harsh Jha, Hamish Ivison, Ian Magnusson, Yizhong Wang, Shane Arora, David Atkinson, Russell Authur, Khyathi Raghavi Chandu, Arman Cohan, Jennifer Dumas, Yanai Elazar, Yuling Gu, Jack Hessel, Tushar Khot, William Merrill, Jacob Morrison, Niklas Muennighoff, Aakanksha Naik, Crystal Nam, Matthew E. Peters, Valentina Pyatkin, Abhilasha Ravichander, Dustin Schwenk, Saurabh Shah, Will Smith, Emma Strubell, Nishant Subramani, Mitchell Wortsman, Pradeep Dasigi, Nathan Lambert, Kyle Richardson, Luke Zettlemoyer, Jesse Dodge, Kyle Lo, Luca Soldaini, Noah A. Smith, and Hannaneh Hajishirzi. 2024. OLMo: Accelerating the Science of Language Models. ArXiv:2402.00838 [cs].

Guerdan, Luke, Amanda Coston, Zhiwei Steven Wu, and Kenneth Holstein. 2023. Ground(less) Truth: A Causal Framework for Proxy Labels in Human-Algorithm Decision-Making. In *2023 ACM Conference on Fairness, Accountability, and Transparency*, pages 688–704, ACM, Chicago IL USA.

Guleria, Ankita, Kewal Krishan, Vishal Sharma, and Tanuj Kanchan. 2023. ChatGPT: ethical concerns and challenges in academics and research. *The Journal of Infection in Developing Countries*, 17(09):1292–1299.

Guo, Yue, Yi Yang, and A. Abbasi. 2022. Auto-Debias: Debiasing Masked Language Models with Automated Biased Prompts. In *Annual Meeting of the Association for Computational Linguistics*.

Gupta, Soumyajit, Sooyong Lee, Maria De-Arteaga, and Matthew Lease. 2023. Same same, but different: Conditional multi-task learning for demographic-specific toxicity detection. In *Proceedings of the ACM Web Conference 2023*, WWW '23, page 3689–3700, Association for Computing Machinery, New York, NY, USA.

Gupta, Vipul, Pranav Narayanan Venkit, Hugo Laurençon, Shomir Wilson, and Rebecca J. Passonneau. 2024. CALM : A Multi-task Benchmark for Comprehensive Assessment of Language Model Bias. ArXiv:2308.12539 [cs].

Guyan, Kevin. 2021. Constructing a queer population? Asking about sexual orientation in Scotland's 2022 census. *Journal of Gender Studies*, 0(0):1–11.

Haque, MD Romael, Devansh Saxena, Katy Weathington, Joseph Chudzik, and Shion Guha. 2024. Are we asking the right questions?: Designing for community stakeholders' interactions with ai in policing. In *Proceedings of the CHI Conference on Human Factors in Computing Systems*, pages 1–20.

Hara, Kotaro, Abigail Adams, Kristy Milland, Saiph Savage, Chris Callison-Burch, and Jeffrey P. Bigham. 2018. A Data-Driven Analysis of Workers' Earnings on Amazon Mechanical Turk.

*Proceedings of the 2018 CHI Conference on Human Factors in Computing Systems*, pages 1–14. Conference Name: CHI '18: CHI Conference on Human Factors in Computing Systems ISBN: 9781450356206 Place: Montreal QC Canada Publisher: ACM.

Haroutunian, Levon. 2022. Ethical Considerations for Low-resourced Machine Translation. In *Proceedings of the 60th Annual Meeting of the Association for Computational Linguistics: Student Research Workshop*, pages 44–54, Association for Computational Linguistics, Dublin, Ireland.

Havens, Lucy, Melissa Terras, Benjamin Bach, and Beatrice Alex. 2020. Situated Data, Situated Systems: A Methodology to Engage with Power Relations in Natural Language Processing Research. *arXiv:2011.05911 [cs]*. ArXiv: 2011.05911.

Heikkilä, Melissa. 2022. Artists can now opt out of the next version of Stable Diffusion. *MIT Technology Review*.

Henderson, Peter, Jie Hu, Joshua Romoff, E. Brunskill, Dan Jurafsky, and Joelle Pineau. 2020. Towards the Systematic Reporting of the Energy and Carbon Footprints of Machine Learning. *ArXiv*.

Hendrycks, Dan, Collin Burns, Steven Basart, Andrew Critch, Jerry Zheng Li, Dawn Xiaodong Song, and Jacob Steinhardt. 2020. Aligning AI With Shared Human Values. *ArXiv*, abs/2008.02275.

Hong, Rachel, William Agnew, Tadayoshi Kohno, and Jamie Morgenstern. 2024. Who's in and who's out? A case study of multimodal CLIP-filtering in DataComp. ArXiv:2405.08209 [cs].

Hooker, Sara. 2021. Moving beyond "algorithmic bias is a data problem". *Patterns*, 2(4):100241.

Hooker, Sara, Nyalleng Moorosi, Gregory Clark, Samy Bengio, and Emily Denton. 2020. Characterising Bias in Compressed Models. ArXiv:2010.03058 [cs].

Hovy, Dirk and Shannon L. Spruit. 2016. The Social Impact of Natural Language Processing. In *Proceedings of the 54th Annual Meeting of the Association for Computational Linguistics (Volume 2: Short Papers)*, pages 591–598, Association for Computational Linguistics, Berlin, Germany.

Huang, Jie, Hanyin Shao, and Kevin Chen-Chuan Chang. 2022. Are Large Pre-Trained Language Models Leaking Your Personal Information? In *Findings of the Association for Computational Linguistics: EMNLP 2022*, pages 2038–2047, Association for Computational Linguistics, Abu Dhabi, United Arab Emirates.

Huang, Olivia, Eve Fleisig, and Dan Klein. 2023. Incorporating worker perspectives into MTurk annotation practices for NLP. In *Proceedings of the 2023 Conference on Empirical Methods in Natural Language Processing*, pages 1010–1028, Association for Computational Linguistics, Singapore.

Huang, Xiaowei, Wenjie Ruan, Wei Huang, Gao Jin, Yizhen Dong, Changshun Wu, Saddek Bensalem, Ronghui Mu, Yi Qi, Xingyu Zhao, Kaiwen Cai, Yanghao Zhang, Sihao Wu, Peipei Xu, Dengyu Wu, André Freitas, and Mustafa A. Mustafa. 2023a. A Survey of Safety and Trustworthiness of Large Language Models through the Lens of Verification and Validation. *ArXiv*, abs/2305.11391.

Huang, Yue, Qihui Zhang, Philip S. Yu, and Lichao Sun. 2023b. TrustGPT: A Benchmark for Trustworthy and Responsible Large Language Models. *ArXiv*, abs/2306.11507.

Inie, Nanna and Leon Derczynski. 2021. An IDR Framework of Opportunities and Barriers between HCI and NLP. In *Proceedings of the First Workshop on Bridging Human–Computer Interaction and Natural Language Processing*, pages 101–108, Association for Computational Linguistics, Online.

Jamieson, Michelle K., Gisela H. Govaart, and Madeleine Pownall. 2023. Reflexivity in quantitative research: A rationale and beginner's guide. *Social and Personality Psychology Compass*, 17(4):e12735.

Jernite, Yacine, Huu Nguyen, Stella Biderman, Anna Rogers, Maraim Masoud, Valentin Danchev, Samson Tan, Alexandra Sasha Luccioni, Nishant Subramani, Gérard Dupont, Jesse Dodge, Kyle Lo, Zeerak Talat, Isaac Johnson, Dragomir Radev, Somaieh Nikpoor, Jörg Frohberg, Aaron Gokaslan, Peter Henderson, Rishi Bommasani, and Margaret Mitchell. 2022. Data Governance in the Age of Large-Scale Data-Driven Language Technology. In *2022 ACM Conference on Fairness, Accountability, and Transparency*, pages 2206–2222. ArXiv:2206.03216 [cs].

Ji, Jiaming, Tianyi Qiu, Boyuan Chen, Borong Zhang, Hantao Lou, Kaile Wang, Yawen Duan, Zhonghao He, Jiayi Zhou, Zhaowei Zhang, et al. 2023. Ai alignment: A comprehensive survey. *arXiv preprint arXiv:2310.19852*.

Jurgens, David, Libby Hemphill, and Eshwar Chandrasekharan. 2019. A Just and Comprehensive Strategy for Using NLP to Address Online Abuse. In *Proceedings of the 57th Annual Meeting of the Association for Computational Linguistics*, pages 3658–3666, Association for Computational Linguistics, Florence, Italy.
Jurgens, David, Yulia Tsvetkov, and Dan Jurafsky. 2017. Incorporating Dialectal Variability for Socially Equitable Language Identification. In *Proceedings of the 55th Annual Meeting of the Association for Computational Linguistics (Volume 2: Short Papers)*, pages 51–57, Association for Computational Linguistics, Vancouver, Canada.
Kaffee, Lucie-Aimée, Arnav Arora, Zeerak Talat, and Isabelle Augenstein. 2023. Thorny Roses: Investigating the Dual Use Dilemma in Natural Language Processing. Publisher: [object Object] Version Number: 3.
Karamolegkou, Antonia, Jiaang Li, Li Zhou, and Anders Søgaard. 2023. Copyright Violations and Large Language Models. In *Proceedings of the 2023 Conference on Empirical Methods in Natural Language Processing*, pages 7403–7412, Association for Computational Linguistics, Singapore.
Kasirzadeh, Atoosa. 2021. Reasons, values, stakeholders: A philosophical framework for explainable artificial intelligence. *arXiv preprint arXiv:2103.00752*.
Kasirzadeh, Atoosa and Iason Gabriel. 2023. In conversation with artificial intelligence: aligning language models with human values. *Philosophy & Technology*, 36(2):27.
Kasirzadeh, Atoosa and Colin Klein. 2021. The ethical gravity thesis: Marrian levels and the persistence of bias in automated decision-making systems. In *Proceedings of the 2021 AAAI/ACM Conference on AI, Ethics, and Society*, pages 618–626.
Kasneci, Enkelejda, Kathrin Seßler, Stefan Küchemann, Maria Bannert, Daryna Dementieva, Frank Fischer, Urs Gasser, Georg Groh, Stephan Günnemann, Eyke Hüllermeier, et al. 2023. Chatgpt for good? on opportunities and challenges of large language models for education. *Learning and individual differences*, 103:102274.
Kaun, Anne and Fredrik Stiernstedt. 2020. Prison media work: from manual labor to the work of being tracked. *Media, Culture & Society*, 42(7-8):1277–1292. Publisher: SAGE Publications Ltd.
Kawakami, Anna, Amanda Coston, Haiyi Zhu, Hoda Heidari, and Kenneth Holstein. 2024. The Situate AI Guidebook: Co-Designing a Toolkit to Support Multi-Stakeholder Early-stage Deliberations Around Public Sector AI Proposals. ArXiv:2402.18774 [cs].
Kekulluoglu, Dilara, Nadin Kokciyan, and Pinar Yolum. 2018. Preserving Privacy as Social Responsibility in Online Social Networks. *ACM Trans. Internet Technol.*, 18(4):42:1–42:22.
Kennedy, Ryan, Scott Clifford, Tyler Burleigh, Philip D Waggoner, Ryan Jewell, and Nicholas JG Winter. 2020. The shape of and solutions to the mturk quality crisis. *Political Science Research and Methods*, 8(4):614–629.
Kim, Siwon, Sangdoo Yun, Hwaran Lee, Martin Gubri, Sungroh Yoon, and Seong Joon Oh. 2023. ProPILE: Probing Privacy Leakage in Large Language Models. *Advances in Neural Information Processing Systems*, 36:20750–20762.
Kirchenbauer, John, Jonas Geiping, Yuxin Wen, Jonathan Katz, Ian Miers, and Tom Goldstein. 2023. A Watermark for Large Language Models. In *International Conference on Machine Learning*.
Kirk, Hannah Rose, Bertie Vidgen, Paul Röttger, and Scott A Hale. 2023. The empty signifier problem: Towards clearer paradigms for operationalising" alignment" in large language models. *arXiv preprint arXiv:2310.02457*.
Klassen, Shamika and Casey Fiesler. 2022. "This Isn't Your Data, Friend": Black Twitter as a Case Study on Research Ethics for Public Data. *Social Media + Society*, 8(4). Publisher: SAGE Publications Ltd.
Knowles, Bran, Jasmine Fledderjohann, John T. Richards, and Kush R. Varshney. 2023. Trustworthy AI and the Logics of Intersectional Resistance. In *Proceedings of the 2023 ACM Conference on Fairness, Accountability, and Transparency*, FAccT '23, pages 172–182, Association for Computing Machinery, New York, NY, USA.
Kumar, Sachin, Vidhisha Balachandran, Lucille Njoo, Antonios Anastasopoulos, and Yulia Tsvetkov. 2022. Language Generation Models Can Cause Harm: So What Can We Do About It? An Actionable Survey. In *Conference of the European Chapter of the Association for Computational Linguistics*.
Lacoste, Alexandre, Alexandra Luccioni, Victor Schmidt, and Thomas Dandres. 2019. Quantifying the carbon emissions of machine learning. *arXiv preprint arXiv:1910.09700*.
LaCroix, Travis and Alexandra Sasha Luccioni. 2022. Metaethical Perspectives on 'Benchmarking' AI Ethics. Publisher: arXiv Version Number: 1.
Langer, Markus, Daniel Oster, Timo Speith, Holger Hermanns, Lena Kästner, Eva Schmidt, Andreas Sesing, and Kevin Baum. 2021. What do we want from Explainable Artificial

Intelligence (XAI)? – A stakeholder perspective on XAI and a conceptual model guiding interdisciplinary XAI research. *Artificial Intelligence*, 296:103473.
Larson, Brian. 2017. Gender as a Variable in Natural-Language Processing: Ethical Considerations. In *Proceedings of the First ACL Workshop on Ethics in Natural Language Processing*, pages 1–11, Association for Computational Linguistics, Valencia, Spain.
Lee, Jooyoung, Thai Le, Jinghui Chen, and Dongwon Lee. 2022. Do Language Models Plagiarize? *Proceedings of the ACM Web Conference 2023*.
Lee, Min Kyung. 2018. Understanding perception of algorithmic decisions: Fairness, trust, and emotion in response to algorithmic management. *Big Data & Society*, 5(1):2053951718756684. Publisher: SAGE Publications Ltd.
Lee, Nayeon, Chani Jung, Junho Myung, Jiho Jin, Jose Camacho-Collados, Juho Kim, and Alice Oh. 2024. Exploring cross-cultural differences in English hate speech annotations: From dataset construction to analysis. In *Proceedings of the 2024 Conference of the North American Chapter of the Association for Computational Linguistics: Human Language Technologies (Volume 1: Long Papers)*, pages 4205–4224, Association for Computational Linguistics, Mexico City, Mexico.
Lee, Noah, Na Min An, and James Thorne. 2023. Can large language models capture dissenting human voices? In *Proceedings of the 2023 Conference on Empirical Methods in Natural Language Processing*, pages 4569–4585, Association for Computational Linguistics, Singapore.
Leidner, Jochen L. and Vassilis Plachouras. 2017. Ethical by Design: Ethics Best Practices for Natural Language Processing. In *Proceedings of the First ACL Workshop on Ethics in Natural Language Processing*, pages 30–40, Association for Computational Linguistics, Valencia, Spain.
Leonardelli, Elisa, Gavin Abercrombie, Dina Almanea, Valerio Basile, Tommaso Fornaciari, Barbara Plank, Verena Rieser, Alexandra Uma, and Massimo Poesio. 2023. SemEval-2023 task 11: Learning with disagreements (LeWiDi). In *Proceedings of the 17th International Workshop on Semantic Evaluation (SemEval-2023)*, pages 2304–2318, Association for Computational Linguistics, Toronto, Canada.
Levy, Sharon, Emily Allaway, Melanie Subbiah, Lydia B. Chilton, Desmond Upton Patton, Kathleen McKeown, and William Yang Wang. 2022. SafeText: A Benchmark for Exploring Physical Safety in Language Models. In *Conference on Empirical Methods in Natural Language Processing*.
Lewis, Dave, Linda Hogan, David Filip, and P. J. Wall. 2020. Global Challenges in the Standardization of Ethics for Trustworthy AI. *Journal of ICT Standardization*.
Liang, Calvin. 2021. Reflexivity, positionality, and disclosure in HCI.
Liang, Calvin A., Sean A. Munson, and Julie A. Kientz. 2021. Embracing Four Tensions in Human-Computer Interaction Research with Marginalized People. *ACM Transactions on Computer-Human Interaction*, 28(2):1–47.
Liang, Paul Pu, Irene Mengze Li, Emily Zheng, Yao Chong Lim, Ruslan Salakhutdinov, and Louis-Philippe Morency. 2020. Towards Debiasing Sentence Representations. In *Proceedings of the 58th Annual Meeting of the Association for Computational Linguistics*, pages 5502–5515, Association for Computational Linguistics, Online.
Liang, Weixin, Mert Yuksekgonul, Yining Mao, Eric Wu, and James Zou. 2023. GPT detectors are biased against non-native English writers. *Patterns*, 4(7):100779.
Lignos, Constantine, Nolan Holley, Chester Palen-Michel, and Jonne Sälevä. 2022. Toward More Meaningful Resources for Lower-resourced Languages. In *Findings of the Association for Computational Linguistics: ACL 2022*, pages 523–532, Association for Computational Linguistics, Dublin, Ireland.
Liu, Jie et al. 2012. The enterprise risk management and the risk oriented internal audit. *Ibusiness*, 4(03):287.
Liu, Zoey, Crystal Richardson, Richard Hatcher, and Emily Prud'hommeaux. 2022. Not always about you: Prioritizing community needs when developing endangered language technology. In *Proceedings of the 60th Annual Meeting of the Association for Computational Linguistics (Volume 1: Long Papers)*, pages 3933–3944, Association for Computational Linguistics, Dublin, Ireland.
Lopez Long, Holly, Alexandra O'Neil, and Sandra Kübler. 2021. On the Interaction between Annotation Quality and Classifier Performance in Abusive Language Detection. In *Proceedings of the International Conference on Recent Advances in Natural Language Processing (RANLP 2021)*, pages 868–875, INCOMA Ltd., Held Online.
Lucchi, Nicola. 2023. Chatgpt: A case study on copyright challenges for generative artificial intelligence systems. *European Journal of Risk Regulation*, page 1–23.

Luccioni, Alexandra and Joseph Viviano. 2021. What's in the Box? An Analysis of Undesirable Content in the Common Crawl Corpus. In *Proceedings of the 59th Annual Meeting of the Association for Computational Linguistics and the 11th International Joint Conference on Natural Language Processing (Volume 2: Short Papers)*, pages 182–189, Association for Computational Linguistics, Online.

Luccioni, Alexandra Sasha and Alex Hernandez-Garcia. 2023. Counting Carbon: A Survey of Factors Influencing the Emissions of Machine Learning. ArXiv:2302.08476 [cs].

Luccioni, Alexandra Sasha, Sylvain Viguier, and Anne-Laure Ligozat. 2023. Estimating the carbon footprint of bloom, a 176b parameter language model. *Journal of Machine Learning Research*, 24(253):1–15.

Luccioni, Sasha, Yacine Jernite, and Emma Strubell. 2024. Power Hungry Processing: Watts Driving the Cost of AI Deployment? In *The 2024 ACM Conference on Fairness, Accountability, and Transparency*, pages 85–99, ACM, Rio de Janeiro Brazil.

Lukas, Nils, A. Salem, Robert Sim, Shruti Tople, Lukas Wutschitz, and Santiago Zanella-B'eguelin. 2023. Analyzing Leakage of Personally Identifiable Information in Language Models. *2023 IEEE Symposium on Security and Privacy (SP)*, pages 346–363.

Lund, Brady D., Ting Wang, Nishith Reddy Mannuru, Bing Nie, Somipam Shimray, and Ziang Wang. 2023. ChatGPT and a new academic reality: Artificial Intelligence-written research papers and the ethics of the large language models in scholarly publishing. *Journal of the Association for Information Science and Technology*, 74(5):570–581.

Madaio, Michael, Lisa Egede, Hariharan Subramonyam, Jennifer Wortman Vaughan, and Hanna Wallach. 2022. Assessing the fairness of ai systems: Ai practitioners' processes, challenges, and needs for support. *Proceedings of the ACM on Human-Computer Interaction*, 6(CSCW1):1–26.

Madnani, Nitin, Anastassia Loukina, Alina von Davier, Jill Burstein, and Aoife Cahill. 2017. Building Better Open-Source Tools to Support Fairness in Automated Scoring. In *Proceedings of the First ACL Workshop on Ethics in Natural Language Processing*, pages 41–52, Association for Computational Linguistics, Valencia, Spain.

Mahelona, Keoni, Gianna Leoni, Suzanne Duncan, and Miles Thompson. 2023. OpenAI's whisper is another case study in colonisation. *Papa Reo*.

Malazita, James W and Korryn Resetar. 2019. Infrastructures of abstraction: how computer science education produces anti-political subjects. *Digital Creativity*, 30(4):300–312.

Mancosu, Moreno and Federico Vegetti. 2020. What You Can Scrape and What Is Right to Scrape: A Proposal for a Tool to Collect Public Facebook Data. *Social Media + Society*, 6(3):2056305120940703. Publisher: SAGE Publications Ltd.

Manerba, Marta Marchiori and Sara Tonelli. 2021. Fine-Grained Fairness Analysis of Abusive Language Detection Systems with CheckList. In *Proceedings of the 5th Workshop on Online Abuse and Harms (WOAH 2021)*, pages 81–91, Association for Computational Linguistics, Online.

Markelius, Alva, Connor Wright, Joahna Kuiper, Natalie Delille, and Yu-Ting Kuo. 2024. The mechanisms of ai hype and its planetary and social costs. *AI and Ethics*, pages 1–16.

Markl, Nina. 2022. Mind the data gap(s): Investigating power in speech and language datasets. In *Proceedings of the Second Workshop on Language Technology for Equality, Diversity and Inclusion*, pages 1–12, Association for Computational Linguistics, Dublin, Ireland.

Maronikolakis, Antonis, Axel Wisiorek, Leah Nann, Haris Jabbar, Sahana Udupa, and Hinrich Schuetze. 2022. Listening to Affected Communities to Define Extreme Speech: Dataset and Experiments. In *Findings of the Association for Computational Linguistics: ACL 2022*, pages 1089–1104, Association for Computational Linguistics, Dublin, Ireland.

Mason, Winter and Siddharth Suri. 2012. Conducting behavioral research on Amazon's Mechanical Turk. *Behavior Research Methods*, 44(1):1–23.

May, Chandler, Alex Wang, Shikha Bordia, Samuel R. Bowman, and Rachel Rudinger. 2019. On Measuring Social Biases in Sentence Encoders. In *Proceedings of the 2019 Conference of the North American Chapter of the Association for Computational Linguistics: Human Language Technologies, Volume 1 (Long and Short Papers)*, pages 622–628, Association for Computational Linguistics, Minneapolis, Minnesota.

McDuff, Daniel, Tim Korjakow, Scott Cambo, Jesse Josua Benjamin, Jenny Lee, Yacine Jernite, Carlos Muñoz Ferrandis, Aaron Gokaslan, Alek Tarkowski, Joseph Lindley, A. Feder Cooper, and Danish Contractor. 2024. On the Standardization of Behavioral Use Clauses and Their Adoption for Responsible Licensing of AI.

McNamara, Andrew, Justin Smith, and Emerson Murphy-Hill. 2018. Does acm's code of ethics change ethical decision making in software development? In *Proceedings of the 2018 26th ACM* *joint meeting on european software engineering conference and symposium on the foundations of software engineering*, pages 729–733.
Meng, Nicole, Dilara Keküllüoğlu, and Kami Vaniea. 2021. Owning and Sharing: Privacy Perceptions of Smart Speaker Users. *Proc. ACM Hum.-Comput. Interact.*, 5(CSCW1):45:1–45:29.
Mieskes, Margot. 2017. A Quantitative Study of Data in the NLP community. In *Proceedings of the First ACL Workshop on Ethics in Natural Language Processing*, pages 23–29, Association for Computational Linguistics, Valencia, Spain.
Miller, Boaz. 2021. Is technology value-neutral? *Science, Technology, & Human Values*, 46(1):53–80.
Min, Bonan, Hayley Ross, Elior Sulem, Amir Pouran Ben Veyseh, Thien Huu Nguyen, Oscar Sainz, Eneko Agirre, Ilana Heintz, and Dan Roth. 2023. Recent Advances in Natural Language Processing via Large Pre-trained Language Models: A Survey. *ACM Comput. Surv.*, 56(2):30:1–30:40.
Mitchell, Margaret, Simone Wu, Andrew Zaldivar, Parker Barnes, Lucy Vasserman, Ben Hutchinson, Elena Spitzer, Inioluwa Deborah Raji, and Timnit Gebru. 2019. Model Cards for Model Reporting. *Proceedings of the Conference on Fairness, Accountability, and Transparency*, pages 220–229. ArXiv: 1810.03993.
Mohammad, Saif. 2022. Ethics Sheets for AI Tasks. In *Proceedings of the 60th Annual Meeting of the Association for Computational Linguistics (Volume 1: Long Papers)*, pages 8368–8379, Association for Computational Linguistics, Dublin, Ireland.
Mokhberian, Negar, Myrl Marmarelis, Frederic Hopp, Valerio Basile, Fred Morstatter, and Kristina Lerman. 2024. Capturing perspectives of crowdsourced annotators in subjective learning tasks. In *Proceedings of the 2024 Conference of the North American Chapter of the Association for Computational Linguistics: Human Language Technologies (Volume 1: Long Papers)*, pages 7337–7349, Association for Computational Linguistics, Mexico City, Mexico.
Mulvin, Dylan. 2021. *Proxies: The Cultural Work of Standing In*. Infrastructures Ser. The MIT Press, Cambridge, Massachusetts.
Munn, Luke. 2022. The uselessness of AI ethics. *AI and Ethics*.
Nalbandian, Lucia. 2022. An eye for an 'I:' a critical assessment of artificial intelligence tools in migration and asylum management. *Comparative Migration Studies*, 10(1):32.
Nathan, Lisa P., Predrag V. Klasnja, and Batya Friedman. 2007. Value scenarios: a technique for envisioning systemic effects of new technologies. In *CHI '07 Extended Abstracts on Human Factors in Computing Systems*, CHI EA '07, pages 2585–2590, Association for Computing Machinery, New York, NY, USA.
Navigli, Roberto, Simone Conia, and Björn Ross. 2023. Biases in large language models: origins, inventory, and discussion. *ACM Journal of Data and Information Quality*, 15(2):1–21.
Niven, Timothy and Hung-Yu Kao. 2019. Probing neural network comprehension of natural language arguments. In *Proceedings of the 57th Annual Meeting of the Association for Computational Linguistics*, pages 4658–4664, Association for Computational Linguistics, Florence, Italy.
OpenAI. 2024. How ChatGPT and our language models are developed. `https://help.openai.com/en/articles/7842364-how-chatgpt-and-our-language-models-are-developed`. Accessed 14 August 2024.
Orgad, Hadas, Seraphina Goldfarb-Tarrant, and Yonatan Belinkov. 2022. How Gender Debiasing Affects Internal Model Representations, and Why It Matters. In *Proceedings of the 2022 Conference of the North American Chapter of the Association for Computational Linguistics: Human Language Technologies*, pages 2602–2628, Association for Computational Linguistics, Seattle, United States.
Ovalle, Anaelia, Ninareh Mehrabi, Palash Goyal, Jwala Dhamala, Kai-Wei Chang, Richard Zemel, Aram Galstyan, and Rahul Gupta. 2023a. Are you talking to ['xem'] or ['x', 'em']? On Tokenization and Addressing Misgendering in LLMs with Pronoun Tokenization Parity. ArXiv:2312.11779 [cs].
Ovalle, Anaelia, Arjun Subramonian, Vagrant Gautam, Gilbert Gee, and Kai-Wei Chang. 2023b. Factoring the Matrix of Domination: A Critical Review and Reimagination of Intersectionality in AI Fairness. In *Proceedings of the 2023 AAAI/ACM Conference on AI, Ethics, and Society*, pages 496–511, ACM, Montr\'{e}al QC Canada.
Pan, Yikang, Liangming Pan, Wenhu Chen, Preslav Nakov, Min-Yen Kan, and William Wang. 2023. On the Risk of Misinformation Pollution with Large Language Models. In *Findings of the Association for Computational Linguistics: EMNLP 2023*, pages 1389–1403, Association for Computational Linguistics, Singapore.
Patterson, David, Joseph Gonzalez, Urs Hölzle, Quoc Le, Chen Liang, Lluis-Miquel Munguia, Daniel Rothchild, David R. So, Maud Texier, and Jeff Dean. 2022. The Carbon Footprint of Machine Learning Training Will Plateau, Then Shrink. *Computer*, 55(7):18–28. Conference Name: Computer.
Paullada, Amandalynne, Inioluwa Deborah Raji, Emily M. Bender, Emily L. Denton, and A. Hanna. 2020. Data and its (dis)contents: A survey of dataset development and use in machine learning research. *Patterns*.
Perez, Ethan, Saffron Huang, Francis Song, Trevor Cai, Roman Ring, John Aslanides, Amelia Glaese, Nathan McAleese, and Geoffrey Irving. 2022. Red Teaming Language Models with Language Models. In *Conference on Empirical Methods in Natural Language Processing*.
Perrigo, Billy. 2023. 150 AI Workers Vote to Unionize at Nairobi Meeting.
Petrov, Aleksandar, Emanuele La Malfa, Philip H. S. Torr, and Adel Bibi. 2023. Language Model Tokenizers Introduce Unfairness Between Languages. *ArXiv*, abs/2305.15425.
Pistilli, Giada, Alina Leidinger, Yacine Jernite, Atoosa Kasirzadeh, Alexandra Sasha Luccioni, and Margaret Mitchell. 2024. Civics: Building a dataset for examining culturally-informed values in large language models. *arXiv preprint arXiv:2405.13974*.
Van de Poel, Ibo. 2020. Embedding values in artificial intelligence (ai) systems. *Minds and machines*, 30(3):385–409.
Rae, Jack W., Sebastian Borgeaud, Trevor Cai, Katie Millican, Jordan Hoffmann, Francis Song, John Aslanides, Sarah Henderson, Roman Ring, Susannah Young, Eliza Rutherford, Tom Hennigan, Jacob Menick, Albin Cassirer, Richard Powell, George van den Driessche, Lisa Anne Hendricks, Maribeth Rauh, Po-Sen Huang, Amelia Glaese, Johannes Welbl, Sumanth Dathathri, Saffron Huang, Jonathan Uesato, John F. J. Mellor, Irina Higgins, Antonia Creswell, Nathan McAleese, Amy Wu, Erich Elsen, Siddhant M. Jayakumar, Elena Buchatskaya, David Budden, Esme Sutherland, Karen Simonyan, Michela Paganini, L. Sifre, Lena Martens, Xiang Lorraine Li, Adhiguna Kuncoro, Aida Nematzadeh, Elena Gribovskaya, Domenic Donato, Angeliki Lazaridou, Arthur Mensch, Jean-Baptiste Lespiau, Maria Tsimpoukelli, N. K. Grigorev, Doug Fritz, Thibault Sottiaux, Mantas Pajarskas, Tobias Pohlen, Zhitao Gong, Daniel Toyama, Cyprien de Masson d'Autume, Yujia Li, Tayfun Terzi, Vladimir Mikulik, Igor Babuschkin, Aidan Clark, Diego de Las Casas, Aurelia Guy, Chris Jones, James Bradbury, Matthew G. Johnson, Blake A. Hechtman, Laura Weidinger, Iason Gabriel, William S. Isaac, Edward Lockhart, Simon Osindero, Laura Rimell, Chris Dyer, Oriol Vinyals, Kareem W. Ayoub, Jeff Stanway, L. L. Bennett, Demis Hassabis, Koray Kavukcuoglu, and Geoffrey Irving. 2021. Scaling Language Models: Methods, Analysis & Insights from Training Gopher. *ArXiv*, abs/2112.11446.
Raji, Inioluwa Deborah, Emily M. Bender, Amandalynne Paullada, Emily Denton, and Alex Hanna. 2021. AI and the Everything in the Whole Wide World Benchmark. ArXiv:2111.15366 [cs].
Raji, Inioluwa Deborah, Morgan Klaus Scheuerman, and Razvan Amironesei. 2021. You can't sit with us: Exclusionary pedagogy in ai ethics education. In *Proceedings of the 2021 ACM conference on fairness, accountability, and transparency*, pages 515–525.
Raji, Inioluwa Deborah, Andrew Smart, Rebecca N White, Margaret Mitchell, Timnit Gebru, Ben Hutchinson, Jamila Smith-Loud, Daniel Theron, and Parker Barnes. 2020. Closing the AI Accountability Gap: Defining an End-to-End Framework for Internal Algorithmic Auditing. page 12.
Ramesh, Krithika, Arnav Chavan, Shrey Pandit, and Sunayana Sitaram. 2023. A Comparative Study on the Impact of Model Compression Techniques on Fairness in Language Models. In *Annual Meeting of the Association for Computational Linguistics*.
Ramesh, Krithika, Sunayana Sitaram, and Monojit Choudhury. 2023. Fairness in Language Models Beyond English: Gaps and Challenges. In *Findings*.
Rastogi, Charvi, Marco Tulio Ribeiro, Nicholas King, and Saleema Amershi. 2023. Supporting Human-AI Collaboration in Auditing LLMs with LLMs. *Proceedings of the 2023 AAAI/ACM Conference on AI, Ethics, and Society*.
Rauh, Maribeth, John Mellor, Jonathan Uesato, Po-Sen Huang, Johannes Welbl, Laura Weidinger, Sumanth Dathathri, Amelia Glaese, Geoffrey Irving, Iason Gabriel, William Isaac, and Lisa Anne Hendricks. 2022. Characteristics of Harmful Text: Towards Rigorous Benchmarking of Language Models. Publisher: [object Object] Version Number: 2.

Reid, Corinne, Clara Calia, Cristóbal Guerra, Liz Grant, Matilda Anderson, Khama Chibwana, Paul Kawale, and Action Amos. 2021. Ethics in global research: Creating a toolkit to support integrity and ethical action throughout the research journey. *Research Ethics*, 17(3):359–374.

Ribeiro, Marco Tulio, Tongshuang Wu, Carlos Guestrin, and Sameer Singh. 2020. Beyond Accuracy: Behavioral Testing of NLP Models with CheckList. In *Proceedings of the 58th Annual Meeting of the Association for Computational Linguistics*, pages 4902–4912, Association for Computational Linguistics, Online.

Rocher, Luc, Julien M. Hendrickx, and Yves-Alexandre de Montjoye. 2019. Estimating the success of re-identifications in incomplete datasets using generative models. *Nature Communications*, 10(1):3069. Number: 1 Publisher: Nature Publishing Group.

Rogers, Anna, Timothy Baldwin, and Kobi Leins. 2021. 'Just What do You Think You're Doing, Dave?' A Checklist for Responsible Data Use in NLP. In *Findings of the Association for Computational Linguistics: EMNLP 2021*, pages 4821–4833, Association for Computational Linguistics, Punta Cana, Dominican Republic.

Rottger, Paul, Bertie Vidgen, Dirk Hovy, and Janet Pierrehumbert. 2022. Two contrasting data annotation paradigms for subjective NLP tasks. In *Proceedings of the 2022 Conference of the North American Chapter of the Association for Computational Linguistics: Human Language Technologies*, pages 175–190, Association for Computational Linguistics, Seattle, United States.

Röttger, Paul, Hannah Kirk, Bertie Vidgen, Giuseppe Attanasio, Federico Bianchi, and Dirk Hovy. 2024. XSTest: A Test Suite for Identifying Exaggerated Safety Behaviours in Large Language Models. In *Proceedings of the 2024 Conference of the North American Chapter of the Association for Computational Linguistics: Human Language Technologies (Volume 1: Long Papers)*, pages 5377–5400, Association for Computational Linguistics, Mexico City, Mexico.

Röttger, Paul, Bertie Vidgen, Dong Nguyen, Zeerak Talat, Helen Margetts, and Janet Pierrehumbert. 2021. HateCheck: Functional Tests for Hate Speech Detection Models. In *Proceedings of the 59th Annual Meeting of the Association for Computational Linguistics and the 11th International Joint Conference on Natural Language Processing (Volume 1: Long Papers)*, pages 41–58, Association for Computational Linguistics, Online.

Santurkar, Shibani, Esin Durmus, Faisal Ladhak, Cinoo Lee, Percy Liang, and Tatsunori Hashimoto. 2023. Whose opinions do language models reflect? In *International Conference on Machine Learning*, pages 29971–30004, PMLR.

Santy, Sebastin, Jenny T. Liang, Ronan Le Bras, Katharina Reinecke, and Maarten Sap. 2023. NLPositionality: Characterizing Design Biases of Datasets and Models. ArXiv:2306.01943 [cs].

Sap, Maarten, Swabha Swayamdipta, Laura Vianna, Xuhui Zhou, Yejin Choi, and Noah A. Smith. 2022. Annotators with attitudes: How annotator beliefs and identities bias toxic language detection. In *Proceedings of the 2022 Conference of the North American Chapter of the Association for Computational Linguistics: Human Language Technologies*, pages 5884–5906, Association for Computational Linguistics, Seattle, United States.

Savoldi, Beatrice, Marco Gaido, Luisa Bentivogli, Matteo Negri, and Marco Turchi. 2021. Gender Bias in Machine Translation. *Transactions of the Association for Computational Linguistics*, 9:845–874. Place: Cambridge, MA Publisher: MIT Press.

Scheuerman, Morgan Klaus, Katy Weathington, Tarun Mugunthan, Emily Denton, and Casey Fiesler. 2023. From Human to Data to Dataset: Mapping the Traceability of Human Subjects in Computer Vision Datasets. *Proceedings of the ACM on Human-Computer Interaction*, 7(CSCW1):1–33.

Schick, Timo, Sahana Udupa, and Hinrich Schütze. 2021. Self-Diagnosis and Self-Debiasing: A Proposal for Reducing Corpus-Based Bias in NLP. *Transactions of the Association for Computational Linguistics*, 9:1408–1424.

Schwartz, Roy, Jesse Dodge, Noah A. Smith, and Oren Etzioni. 2019. Green AI.

Selbst, Andrew D., Danah Boyd, Sorelle A. Friedler, Suresh Venkatasubramanian, and Janet Vertesi. 2019. Fairness and Abstraction in Sociotechnical Systems. In *Proceedings of the Conference on Fairness, Accountability, and Transparency*, FAT* '19, pages 59–68, Association for Computing Machinery, New York, NY, USA.

Shin, Philip Wootaek, Jihyun Janice Ahn, Wenpeng Yin, Jack Sampson, and Vijaykrishnan Narayanan. 2024. Can Prompt Modifiers Control Bias? A Comparative Analysis of Text-to-Image Generative Models. ArXiv:2406.05602 [cs] version: 1.

Shmueli, Boaz, Jan Fell, Soumya Ray, and Lun-Wei Ku. 2021. Beyond Fair Pay: Ethical Implications of NLP Crowdsourcing. In *Proceedings of the 2021 Conference of the North American Chapter of the Association for Computational Linguistics: Human Language Technologies*, pages 3758–3769, Association for Computational Linguistics, Online.
Shneiderman, Ben. 2020. Human-Centered Artificial Intelligence: Reliable, Safe & Trustworthy. *International Journal of Human–Computer Interaction*, 36(6):495–504. Publisher: Taylor & Francis _eprint: https://doi.org/10.1080/10447318.2020.1741118.
Sigurgeirsson, Atli and Eddie L. Ungless. 2024. Just Because We Camp, Doesn't Mean We Should: The Ethics of Modelling Queer Voices. ArXiv:2406.07504 [cs].
Sloane, Mona, Emanuel Moss, Olaitan Awomolo, and Laura Forlano. 2022. Participation Is not a Design Fix for Machine Learning. In *Equity and Access in Algorithms, Mechanisms, and Optimization*, pages 1–6, ACM, Arlington VA USA.
Smart, Andrew, Ding Wang, Ellis Monk, Mark Díaz, Atoosa Kasirzadeh, Erin Van Liemt, and Sonja Schmer-Galunder. 2024. Discipline and Label: A WEIRD Genealogy and Social Theory of Data Annotation. ArXiv:2402.06811 [cs].
Smith, Eric Michael, Melissa Hall Melanie Kambadur, Eleonora Presani, and Adina Williams. 2022. "I'm sorry to hear that": finding bias in language models with a holistic descriptor dataset. Technical Report arXiv:2205.09209, arXiv. ArXiv:2205.09209 [cs] version: 1 type: article.
Sok, Sarin and Kimkong Heng. 2023. Chatgpt for education and research: A review of benefits and risks. *Cambodian Journal of Educational Research*, 3(1):110–121.
Solaiman, Irene, Miles Brundage, Jack Clark, Amanda Askell, Ariel Herbert-Voss, Jeff Wu, Alec Radford, Gretchen Krueger, Jong Wook Kim, Sarah Kreps, Miles McCain, Alex Newhouse, Jason Blazakis, Kris McGuffie, and Jasmine Wang. 2019. Release Strategies and the Social Impacts of Language Models. Technical Report arXiv:1908.09203, arXiv. ArXiv:1908.09203 [cs] type: article.
Solaiman, Irene, Zeerak Talat, William Agnew, Lama Ahmad, Dylan Baker, Su Lin Blodgett, Canyu Chen, Hal Daumé III, Jesse Dodge, Isabella Duan, Ellie Evans, Felix Friedrich, Avijit Ghosh, Usman Gohar, Sara Hooker, Yacine Jernite, Ria Kalluri, Alberto Lusoli, Alina Leidinger, Michelle Lin, Xiuzhu Lin, Sasha Luccioni, Jennifer Mickel, Margaret Mitchell, Jessica Newman, Anaelia Ovalle, Marie-Therese Png, Shubham Singh, Andrew Strait, Lukas Struppek, and Arjun Subramonian. 2024. Evaluating the Social Impact of Generative AI Systems in Systems and Society.
Sorensen, Taylor, Jared Moore, Jillian Fisher, Mitchell Gordon, Niloofar Mireshghallah, Christopher Michael Rytting, Andre Ye, Liwei Jiang, Ximing Lu, Nouha Dziri, et al. 2024. A roadmap to pluralistic alignment. *arXiv preprint arXiv:2402.05070*.
Spirling, Arthur. 2023. Why open-source generative AI models are an ethical way forward for science. *Nature*, 616:413.
Steed, Ryan, Swetasudha Panda, Ari Kobren, and Michael L. Wick. 2022. Upstream Mitigation Is Not All You Need: Testing the Bias Transfer Hypothesis in Pre-Trained Language Models. In *ACL*.
Strubell, Emma, Ananya Ganesh, and Andrew McCallum. 2019. Energy and Policy Considerations for Deep Learning in NLP. *Proceedings of the 57th Annual Meeting of the Association for Computational Linguistics*, pages 3645–3650. Conference Name: Proceedings of the 57th Annual Meeting of the Association for Computational Linguistics Place: Florence, Italy Publisher: Association for Computational Linguistics.
Subramani, Nishant, Sasha Luccioni, Jesse Dodge, and Margaret Mitchell. 2023. Detecting Personal Information in Training Corpora: an Analysis. In *Proceedings of the 3rd Workshop on Trustworthy Natural Language Processing (TrustNLP 2023)*, pages 208–220, Association for Computational Linguistics, Toronto, Canada.
Subramanian, Shivashankar, Afshin Rahimi, Timothy Baldwin, Trevor Cohn, and Lea Frermann. 2021. Fairness-aware Class Imbalanced Learning. In *Proceedings of the 2021 Conference on Empirical Methods in Natural Language Processing*, pages 2045–2051, Association for Computational Linguistics, Online and Punta Cana, Dominican Republic.
Sun, Hao, Zhexin Zhang, Jiawen Deng, Jiale Cheng, and Minlie Huang. 2023. Safety Assessment of Chinese Large Language Models. *ArXiv*, abs/2304.10436.
Sun, Tianxiang, Junliang He, Xipeng Qiu, and Xuanjing Huang. 2022. BERTScore is Unfair: On Social Bias in Language Model-Based Metrics for Text Generation. *ArXiv*, abs/2210.07626.
Suresh, Harini and John Guttag. 2021. A Framework for Understanding Sources of Harm throughout the Machine Learning Life Cycle. In *Equity and Access in Algorithms, Mechanisms, and Optimization*, EAAMO '21, pages 1–9, Association for Computing Machinery, New York, NY, USA.

Talat, Zeerak. 2016. Are You a Racist or Am I Seeing Things? Annotator Influence on Hate Speech Detection on Twitter. In *Proceedings of the First Workshop on NLP and Computational Social Science*, pages 138–142, Association for Computational Linguistics, Austin, Texas.
Talat, Zeerak. 2021. *"It Ain't All Good:" Machinic Abuse Detection and Marginalisation in Machine Learning*. Ph.D. thesis, University of Sheffield, Sheffield.
Talat, Zeerak, Thomas Davidson, Dana Warmsley, and Ingmar Weber. 2017. Understanding Abuse: A Typology of Abusive Language Detection Subtasks. In *Proceedings of the First Workshop on Abusive Language Online*, pages 78–84, Association for Computational Linguistics, Vancouver, BC, Canada.
Talat, Zeerak, Smarika Lulz, Joachim Bingel, and Isabelle Augenstein. 2021. Disembodied Machine Learning: On the Illusion of Objectivity in NLP. *arXiv:2101.11974 [cs]*. ArXiv: 2101.11974.
Talat, Zeerak, Aurélie Névéol, Stella Biderman, Miruna Clinciu, Manan Dey, Shayne Longpre, Sasha Luccioni, Maraim Masoud, Margaret Mitchell, Dragomir Radev, Shanya Sharma, Arjun Subramonian, Jaesung Tae, Samson Tan, Deepak Tunuguntla, and Oskar Van Der Wal. 2022. You reap what you sow: On the challenges of bias evaluation under multilingual settings. In *Proceedings of BigScience Episode #5 – Workshop on Challenges & Perspectives in Creating Large Language Models*, pages 26–41, Association for Computational Linguistics, virtual+Dublin.
Tan, Samson, Shafiq Joty, Kathy Baxter, Araz Taeihagh, Gregory A. Bennett, and Min-Yen Kan. 2021. Reliability Testing for Natural Language Processing Systems. In *Proceedings of the 59th Annual Meeting of the Association for Computational Linguistics and the 11th International Joint Conference on Natural Language Processing (Volume 1: Long Papers)*, pages 4153–4169, Association for Computational Linguistics, Online.
Tan, Samson, Shafiq Joty, Lav Varshney, and Min-Yen Kan. 2020. Mind Your Inflections! Improving NLP for Non-Standard Englishes with Base-Inflection Encoding. *Proceedings of the 2020 Conference on Empirical Methods in Natural Language Processing (EMNLP)*, pages 5647–5663. Conference Name: Proceedings of the 2020 Conference on Empirical Methods in Natural Language Processing (EMNLP) Place: Online Publisher: Association for Computational Linguistics.
Tan, Yi Chern and L. Elisa Celis. 2019. Assessing social and intersectional biases in contextualized word representations. In *Proceedings of the 33rd International Conference on Neural Information Processing Systems*, 1185. Curran Associates Inc., Red Hook, NY, USA, pages 13230–13241.
Tatman, Rachael. 2017. Gender and Dialect Bias in YouTube's Automatic Captions. In *Proceedings of the First ACL Workshop on Ethics in Natural Language Processing*, pages 53–59, Association for Computational Linguistics, Valencia, Spain.
Thylstrup, Nanna and Zeerak Talat. 2020. Detecting 'Dirt' and 'Toxicity': Rethinking Content Moderation as Pollution Behaviour.
Toxtli, Carlos, Siddharth Suri, and Saiph Savage. 2021. Quantifying the Invisible Labor in Crowd Work. *Proceedings of the ACM on Human-Computer Interaction*, 5(CSCW2):1–26.
Uma, Alexandra, Dina Almanea, and Massimo Poesio. 2022. Scaling and disagreements: Bias, noise, and ambiguity. *Frontiers in Artificial Intelligence*, 5:818451.
Uma, Alexandra N, Tommaso Fornaciari, Dirk Hovy, Silviu Paun, Barbara Plank, and Massimo Poesio. 2021. Learning from disagreement: A survey. *Journal of Artificial Intelligence Research*, 72:1385–1470.
Ung, Megan, Jing Xu, and Y.-Lan Boureau. 2022. SaFeRDialogues: Taking Feedback Gracefully after Conversational Safety Failures. *undefined*.
Ungless, Eddie, Björn Ross, and Anne Lauscher. 2023. Stereotypes and Smut: The (Mis)representation of Non-cisgender Identities by Text-to-Image Models. In *Findings of the Association for Computational Linguistics: ACL 2023*, Association for Computational Linguistics (ACL).
Ungless, Eddie L., Amy Rafferty, Hrichika Nag, and Björn Ross. 2022. A Robust Bias Mitigation Procedure Based on the Stereotype Content Model. In *Proceedings of the Fifth Workshop on Natural Language Processing and Computational Social Science (NLP+CSS)*, pages 207–217, Association for Computational Linguistics, Abu Dhabi, UAE.
Uzun, Levent. 2023. Are Concerns Related to Artificial Intelligence Development and Use Really Necessary: A Philosophical Discussion. *Digital Society*, 2(3):40.
Vynck, Gerrit De and Nitasha Tiku. 2024. Google takes down Gemini AI image generator. Here's what you need to know. *Washington Post*.

Walczak, Krzysztof and Wojciech Cellary. 2023. Challenges for higher education in the era of widespread access to generative ai. *Economics and Business Review*, 9(2):71–100.
Walter, Maggie, Raymond Lovett, Bobby Maher, Bhiamie Williamson, Jacob Prehn, Gawaian Bodkin-Andrews, and Vanessa Lee. 2021. Indigenous Data Sovereignty in the Era of Big Data and Open Data. *Australian Journal of Social Issues*, 56(2):143–156. _eprint: https://onlinelibrary.wiley.com/doi/pdf/10.1002/ajs4.141.
Wang, Ruotong, F. Maxwell Harper, and Haiyi Zhu. 2020. Factors Influencing Perceived Fairness in Algorithmic Decision-Making: Algorithm Outcomes, Development Procedures, and Individual Differences. In *Proceedings of the 2020 CHI Conference on Human Factors in Computing Systems*, pages 1–14, ACM, Honolulu HI USA.
Wang, Xiaorong, Clara Na, Emma Strubell, Sorelle Friedler, and Sasha Luccioni. 2023. Energy and Carbon Considerations of Fine-Tuning BERT. ArXiv:2311.10267 [cs].
Wei, Mengyi and Zhixuan Zhou. 2022. AI Ethics Issues in Real World: Evidence from AI Incident Database. Publisher: [object Object] Version Number: 2.
Weidinger, Laura, John F. J. Mellor, M. Rauh, C. Griffin, J. Uesato, Po-Sen Huang, M. Cheng, Mia Glaese, B. Balle, A. Kasirzadeh, Z. Kenton, S. Brown, W. Hawkins, T. Stepleton, C. Biles, A. Birhane, Julia Haas, Laura Rimell, Lisa Anne Hendricks, William S. Isaac, Sean Legassick, Geoffrey Irving, and Iason Gabriel. 2021. Ethical and social risks of harm from Language Models. *undefined*.
Weidinger, Laura, Maribeth Rauh, Nahema Marchal, Arianna Manzini, Lisa Anne Hendricks, Juan Mateos-Garcia, Stevie Bergman, Jackie Kay, Conor Griffin, Ben Bariach, et al. 2023. Sociotechnical safety evaluation of generative ai systems. *arXiv preprint arXiv:2310.11986*.
Weidinger, Laura, Jonathan Uesato, Maribeth Rauh, Conor Griffin, Po-Sen Huang, John Mellor, Amelia Glaese, Myra Cheng, Borja Balle, Atoosa Kasirzadeh, et al. 2022. Taxonomy of risks posed by language models. In *Proceedings of the 2022 ACM Conference on Fairness, Accountability, and Transparency*, pages 214–229.
Weisz, Justin D., Michael Muller, Jessica He, and Stephanie Houde. 2023. Toward General Design Principles for Generative AI Applications. Publisher: arXiv Version Number: 1.
Welbl, Johannes, Amelia Glaese, Jonathan Uesato, Sumanth Dathathri, John Mellor, Lisa Anne Hendricks, Kirsty Anderson, Pushmeet Kohli, Ben Coppin, and Po-Sen Huang. 2021. Challenges in Detoxifying Language Models. *arXiv:2109.07445 [cs]*. ArXiv: 2109.07445.
Welch, Charles, Jonathan K. Kummerfeld, Verónica Pérez-Rosas, and Rada Mihalcea. 2020. Compositional demographic word embeddings. In *Proceedings of the 2020 Conference on Empirical Methods in Natural Language Processing (EMNLP)*, pages 4076–4089, Association for Computational Linguistics, Online.
West, Sarah Myers, Meredith Whittaker, and Kate Crawford. 2019. Discriminating Systems: Gender, Race and Power in AI. Technical report, AI Now Institute.
WGA Negotiating Comittee. 2023. WGA on Strike.
Widder, David Gray. 2024. Epistemic Power in AI Ethics Labor: Legitimizing Located Complaints.
Widder, David Gray and Dawn Nafus. 2023. Dislocated accountabilities in the "ai supply chain": Modularity and developers' notions of responsibility. *Big Data & Society*, 10(1):20539517231177620.
Widder, David Gray, Sarah West, and Meredith Whittaker. 2023. Open (For Business): Big Tech, Concentrated Power, and the Political Economy of Open AI.
Winner, Langdon. 1980. Do Artifacts Have Politics? *Daedalus*, 109(1):121–136. Publisher: The MIT Press.
Winograd, Amy. 2022. Loose-Lipped Large Language Models Spill Your Secrets: The Privacy Implications of Large Language Models Notes. *Harvard Journal of Law & Technology (Harvard JOLT)*, 36(2):615–656.
Wittenberg, Chloe, Ziv Epstein, Adam J. Berinsky, and David G. Rand. 2024. Labeling AI-Generated Content: Promises, Perils, and Future Directions. *An MIT Exploration of Generative AI*. Publisher: MIT.
Wong, Richmond Y., Michael A. Madaio, and Nick Merrill. 2023. Seeing Like a Toolkit: How Toolkits Envision the Work of AI Ethics. *Proceedings of the ACM on Human-Computer Interaction*, 7(CSCW1):1–27.
Xiao, Geoffrey. 2020. Bad Bots: Regulating the Scraping of Public Personal Information Notes. *Harvard Journal of Law & Technology (Harvard JOLT)*, 34(2):701–732.

Xu, Albert, Eshaan Pathak, Eric Wallace, Suchin Gururangan, Maarten Sap, and Dan Klein. 2021. Detoxifying Language Models Risks Marginalizing Minority Voices. In *Proceedings of the 2021 Conference of the North American Chapter of the Association for Computational Linguistics: Human Language Technologies*, pages 2390–2397, Association for Computational Linguistics, Online.
Xu, Guangxuan and Qingyuan Hu. 2022. Can Model Compression Improve NLP Fairness. *ArXiv*, abs/2201.08542.
Xu, Ziwei, Sanjay Jain, and Mohan Kankanhalli. 2024. Hallucination is inevitable: An innate limitation of large language models. *arXiv preprint arXiv:2401.11817*.
Yang, Jiancheng, Hongwei Bran Li, and Donglai Wei. 2023. The impact of chatgpt and llms on medical imaging stakeholders: perspectives and use cases. *Meta-Radiology*, page 100007.
Yao, Yifan, Jinhao Duan, Kaidi Xu, Yuanfang Cai, Zhibo Sun, and Yue Zhang. 2024. A survey on large language model (llm) security and privacy: The good, the bad, and the ugly. *High-Confidence Computing*, page 100211.
Zhang, Jizhi, Keqin Bao, Yang Zhang, Wenjie Wang, Fuli Feng, and Xiangnan He. 2023. Is ChatGPT Fair for Recommendation? Evaluating Fairness in Large Language Model Recommendation. *Proceedings of the 17th ACM Conference on Recommender Systems*.
Zhao, Jieyu, Daniel Khashabi, Tushar Khot, Ashish Sabharwal, and Kai-Wei Chang. 2021. Ethical-Advice Taker: Do Language Models Understand Natural Language Interventions? In *Findings of the Association for Computational Linguistics: ACL-IJCNLP 2021*, pages 4158–4164, Association for Computational Linguistics, Online.
Zhuo, Terry Yue, Yujin Huang, Chunyang Chen, and Zhenchang Xing. 2023. Red teaming ChatGPT via Jailbreaking: Bias, Robustness, Reliability and Toxicity. ArXiv:2301.12867 [cs].

## 1. Literature Review Methodology

A primary goal of ETHICS WHITEPAPER is to provide a comprehensive directory of resources for ethical research related to LLMs. As such, a systematic literature review of the ACL Anthology was conducted.[15] The Anthology was searched for papers containing at least one term from the following key term lists in the abstract: related to the type of resource = `tool[a-z]*`, `toolkit`, `[A-za-z]*sheets*`, `guidelines*`, `principles`, `framework`, `approach`; related to ethics = `ethic[a-z]*`, `harms*`, `fair[a-z]*`, `risks*`. These lists were determined by first using more comprehensive lists then eliminating terms to improve the precision of the search. The resulting papers were manually reviewed to determine which were relevant to the scope of ETHICS WHITEPAPER. During the search we identified a 2023 EACL tutorial titled "Understanding Ethics in NLP Authoring and Reviewing" (Benotti et al. 2023). The references for this tutorial were manually reviewed and where relevant included in ETHICS WHITEPAPER.

A second literature review was conducted using Semantic Scholar using the search terms: `toolkit OR sheets OR guideline OR principles OR framework OR approach ethics OR ethical OR harms OR fair OR fairness OR risk AND "language models"`. These were likewise manually reviewed for inclusion.

The resulting resources were categorised by their relevance to different stages in a project's lifespan (from ideation to deployment). Primary themes in the literature were identified and used to structure the section.

As ETHICS WHITEPAPER progressed, additional resources familiar to the authors were added *ad hoc*. Additionally, papers identified during the research review were removed if deemed no longer relevant - thus ETHICS WHITEPAPER does not represent our

15 Note that ETHICS WHITEPAPER is not itself strictly a systematic literature review as resources were added and removed subject to co-author evaluation.

systematic literature review in its entirety, but rather is primarily intended as a practical resource for conducting ethical research related to LLMs. Combining a literature review with our own expertise ensures broad coverage whilst maintaining a pragmatic focus.